\documentclass[fleqn]{iopart}

\usepackage{times}
\usepackage{graphicx,iopams,setstack}

\font\bfgreek=cmmib10
\def\bbu{{\hbox{\bfgreek\char'165}}}
\def\bbw{{\hbox{\bfgreek\char'167}}}
\def\bbs{{\hbox{\bfgreek\char'163}}}
\def\bbf{{\hbox{\bfgreek\char'146}}}
\def\rmO{{\rm O}}
\def\rmd{{\rm d}}

\def\bea{\begin{eqnarray}}
\def\eea{\end{eqnarray}}

\begin{document}

\title[Relativistic MHD and black hole excision]%
{Relativistic MHD and black hole excision: Formulation and initial tests}

\author{David Neilsen, Eric W Hirschmann, and R Steven Millward}
\address{
Department of Physics and Astronomy, 
Brigham Young University, Provo, UT, 84602
}
\ead{david.neilsen@byu.edu}

%
%

\begin{abstract}
A new algorithm for solving the general relativistic MHD equations is
described in this paper.  We design our scheme to incorporate black hole
excision with smooth boundaries, and to simplify solving the combined
Einstein and MHD equations with AMR.  The fluid equations are solved using
a finite difference Convex ENO method.  Excision is implemented using
overlapping grids.  Elliptic and hyperbolic divergence cleaning
techniques allow for maximum flexibility in choosing coordinate
systems, and we compare both methods for a standard problem.
Numerical results of standard test problems are presented in two-dimensional 
flat space using excision, overlapping grids, and elliptic and hyperbolic
divergence cleaning.
\end{abstract}

\submitto{\CQG}

%
%
\section{Introduction}
\label{sec:introduction}

The interaction of strong gravitational and magnetic fields is 
important in a variety of astrophysical phenomena.  The Blandford--Znajek
process~\cite{Blandford:1977}, 
in which a magnetized plasma can extract spin energy
from a black hole, is a promising mechanism for understanding
relativistic jets in AGNs, galactic microquasars, and gamma-ray bursts.  
Furthermore, the magnetorotational instability is an important
mechanism to transport angular momentum in accretion disks~\cite{Balbus}.
Neutron stars and pulsars may also have intense magnetic fields, with magnetars
being an extreme example.  Magnetars are models for soft gamma-ray 
repeaters and anomalous x-ray pulsars.

A large body of work in relativistic magneto-hydrodynamics has
been done in flat space or special relativity.  The incorporation of
significant gravitational effects, however, requires general relativity.
The Arnowitt--Deser--Misner (ADM) formulation of the Einstein equations
is adapted to solving the initial value problem in general 
relativity~\cite{ADM},
and pioneering work in GRMHD using the ADM formulation was done by
Sloan and Smarr~\cite{Sloan:1985}, and Evans and Hawley~\cite{Evans:1988}.
A renewed interest in GRMHD is evident from recent work on both
fixed backgrounds~\cite{Koide:2000,DeVilliers:2002ab,Gammie:2003rj,
Baumgarte:2002vu,Komissarov:2004,Anton:2005gi} and with dynamic 
geometries~\cite{Duez:2005sf,Shibata:2005gp}.

In this paper we describe our method for solving the relativistic
MHD equations.  The equations are derived for a general, arbitrary spacetime.
The standard flat space test
problems are performed on multiple grids with an excised region.
We investigate hyperbolic and elliptic divergence cleaning for the
MHD equations.
Although  the numerical results shown here are
done in flat space, we develop techniques applicable for black hole
spacetimes with excision.
Secondly, we have designed our method for seamless integration with the 
Einstein equations when using adaptive mesh refinement (AMR).  The combination
of general relativity and MHD in evolutions with AMR will be presented
in subsequent papers.

The fluid equations are fundamentally conservation equations, and this 
conservation property can be expressed numerically by using
{\em finite volume} (FV) discretizations. Here the domain is discretized into
volume elements or cells of finite size, and the evolution variables represent
cell averages, such as energy or momentum.   
Essentially Non-Oscillatory (ENO) numerical schemes~\cite{ShuReview}, 
however, allow either
a FV, or, for uniform grids, a {\em finite difference} (FD) discretization.   
FD functions represent point values at discrete points as opposed to 
the averages in FV schemes.
A FD formulation of the MHD equations in general relativity is
compelling for two reasons: (1) communications between arbitrary grids are
simplified as only point values are required; and (2) solving the combined
fluid and Einstein equations with AMR is simplified when both sets of
equations are discretized in the same manner.
The simultaneous refinement of both vertex centered (FD) and
cell centered (FV) grids results in staggered grids.
As the Einstein equations are frequently discretized with finite
differences, a combined general relativity and MHD code can be
simplified if both sets of equations are discretized in the same manner.
We choose a central Convex ENO (CENO) method with FD discretization for the 
MHD equations.

Black hole excision is commonly used in numerical studies of black
hole spacetimes.  The inner excision boundary must be 
carefully constructed such that numerical modes do not enter the
domain through the inner 
boundary~\cite{Scheel:2000,Lehner:2004,Calabrese:2003vy}.
Coordinate systems adapted to the
horizon's geometry allow one to excise the largest volume of spacetime.
This is advantageous because gradients in the gravitational fields become
larger the closer one is to the singularity, requiring greater computational
resources to adequately resolve them.  
Finding a global coordinate system adapted
to all boundaries may only be possible for the most symmetric cases.  Thus,
multiple coordinate patches may be necessary to cover the entire domain.

Different coordinate systems can be implemented computationally using multiple
grids with appropriate communication defined between grids.  One
approach is to use touching or abutting grids, in which all boundary
points on neighboring grids coincide.  These grids have been used in
black hole evolutions with both spectral~\cite{Kidder:2000yq}  
and finite difference methods~\cite{Lehner:2005bz}.  A second approach
uses multiple grids that overlap~\cite{Starius, Henshaw}.  
Information between grids is communicated
via interpolation.  Overlapping grids allow for greater freedom in choosing
numerical schemes, coordinate systems, and moving some grids with respect
to others.
The feasibility of using overlapping grids for moving black holes was
explored by successfully solving the Klein--Gordon equation on fixed black hole
backgrounds for both highly boosted ($v = 0.98c$)
Schwarzschild~\cite{Calabrese:2003vy} and Kerr black ($a=0.99$)
holes~\cite{Calabrese:2004gs}.
It is most natural to implement HRSC schemes for fluids on overlapping
grids.

The magnetic field $\bf B$ is evolved with the MHD fluid equations  and
must also satisfy the constraint $\nabla\cdot {\bf B}=0$.
Experience has shown that error in this constraint will grow
to unacceptable levels (see, for example, 
\fref{Fig:div_cleaning} and discussion below),
leading to unphysical solutions, unless the constraint
is actively enforced.  Moreover, the constraint growth is exacerbated when
weakly hyperbolic formulations of the MHD equations are 
used~\cite{LehnerReula}.
Some techniques to enforce this constraint
include constrained transport, projection
methods, and hyperbolic divergence cleaning~\cite{Toth,Balsara:2003ui}.  
Constrained transport differences the evolution
equations for $\bf B$ such that $\nabla\cdot{\bf B}=0$ is 
satisfied to machine precision for {\it one} particular discrete 
divergence operator. Naturally, the continuum constraint is
only satisfied to the level of truncation error, which can be easily 
seen by evaluating the divergence with a  different, consistent 
discrete divergence operator~\cite{Toth}.  While some constrained transport
schemes may be used with structured AMR~\cite{Balsara:2001rw}, we turn to 
other methods for greater freedom in choosing multiple coordinate systems.
Hyperbolic divergence cleaning
adds a new field designed such that divergence errors are propagated off
the grid~\cite{Dedner:2002}, and is similar to the $\lambda$-system of 
Brodbeck \etal\ for the Einstein equations~\cite{Brodbeck:1998az}.  
Projection methods involve solving an elliptic equation for a correction
to $\bf B$, such that it satisfies the constraint.
Both the projection
method and hyperbolic divergence cleaning are easily implemented with 
overlapping grids.

This paper presents details of our method and gives results of numerical
tests.  The MHD equations in general relativity are derived  in 
\sref{sec:equations}.  \Sref{sec:numerical} presents the
numerical scheme.  
\Sref{sec:divB} discusses divergence cleaning for the MHD equations.
All numerical results are performed on overlapping grids, and are presented
in \Sref{sec:tests}.

%
%
\section{The MHD equations in general relativity}
\label{sec:equations}

We first derive the equations of motion for relativistic MHD and a dynamic 
spacetime. 
The equations are written in conservation form as required for High-Resolution
Shock-Capturing (HRSC) numerical methods.
We then discuss the transformation between conserved and primitive variables.

\subsection{Equations of motion}
\label{subsec:eom}

To begin, we assume a stress energy tensor of the form
\begin{equation} 
T_{ab}  =   \left[ \rho_0 \left( 1 + \epsilon \right) + P \right] u_a u_b
             + P g_{ab}
             + F_{ac} F_{b}^{c}
             - \frac{1}{4} g_{ab} F_{cd} F^{cd} ,
\end{equation} 
where the first few terms describe the fluid and the final two terms the
electromagnetic field.  The fluid and electromagnetic components are coupled
through the relativistic form of Ohm's law:
\begin{equation}
J_a + \left( u_b J^b \right) \,  u_a = \sigma F_{ab} u^b,
\end{equation} 
where $J_a$ is the 4-current.  The ideal MHD
approximation is simply the statement that the fluid has perfect conductivity, 
{\it i.e.}, $\sigma \rightarrow \infty$.
Equivalently, this can be expressed as
\begin{equation} 
F_{ab} \, u^b = 0,
\end{equation} 
which states that the electric field in the frame of the fluid vanishes.  This 
is sometimes referred to as the ``freezing-in" condition of the
magnetic field; namely, in the frame of the fluid, the magnetic field lines
are frozen to the fluid and carried along with it.

With this in mind, a convenient set of substitutions for the electromagnetic 
variables is to define 4-covariant ``electric" and ``magnetic" four-vectors
\begin{equation}
e^a  =  F^{ab} u_b,  \qquad b^a  = *F^{ab} u_b,
\end{equation}
where $*F^{ab} \equiv \epsilon^{abcd} F_{cd} / 2$ and $\epsilon^{abcd}$ is the
standard totally antisymmetric Levi-Civita tensor.
Note that we can write these as
\begin{equation}
 F_{cd} = u_c e_d - u_d e_c - \epsilon_{cdef} u^e b^f, \qquad 
*F_{cd} = u_c b_d - u_d b_c + \epsilon_{cdef} u^e e^f, 
\end{equation}
where we have the constraints $u_a e^a = 0 = u_a b^a$.  All the
information in the
Maxwell tensor, $F_{ab}$, is now contained in these two four vectors.

With these substitutions, the electromagnetic part of the stress tensor
can be written as
\begin{eqnarray}
\fl T_{ab}^{\rm EM}  =    u_a u_b \left[ e_c e^c + b_c b^c \right]
                        + \frac{1}{2} \, g_{ab} \left[ e_c e^c + b_c b^c \right]
                        - e_a e_b - b_a b_b
                        + 2 u_{(a} \epsilon_{b)cde} e^c u^d b^e .
\end{eqnarray}
In the MHD approximation, the electric four vector is identically
zero and the full stress tensor for MHD can be written as
\begin{equation}
T_{ab}  =  \left[ \rho_0 \left( 1 + \epsilon \right) + P + b_c b^c \right] u_a u_b
         + \left[ P + {1\over2} \, b_c b^c \right] \, g_{ab}
            - b_a b_b .
\end{equation}
The matter equations of motion can now be written in conservation
form
\begin{equation}
\nabla_a T^{ab}  =  0, \qquad
\nabla_a {*F}^{ab}  =  0. 
\end{equation}
To these must be appended the baryon conservation equation
$\nabla_a \left( \rho_0 u^a \right) = 0$.  

In a general spacetime we decompose these equations in the
usual ADM 3+1 split by projecting along and orthogonal to a unit normal vector,
$n^a$, which is orthogonal to a foliation of spatial hypersurfaces.
The projection tensor is
\begin{equation}
h_{ab} = g_{ab} + n_a n_b,
\end{equation}
with $g_{ab}$ the metric on the 4-manifold.  The Einstein equations have
the usual 3+1 form with both evolution and constraint equations.  
Because our focus in this paper is developing a robust MHD code, we will emphasize 
and solve the flat spacetime equations in later sections.  However, our approach
in deriving the equations in this section is completely general.  

Conservative variables are defined in the conventional way
\begin{eqnarray}
E     &=&  T_{ab} \, n^a n^b,\\
S_b   &=&  - T_{ac} \, n^a h_b{}^c,\\
\left( \perp\! T \right)_{cd}   &=&  T_{ab} \, h^a{}_c h^b{}_d.
\end{eqnarray}
With respect to the MHD stress tensor, these give 
\begin{eqnarray}
\fl E   &=&
  \left[ \rho_0 \left( 1 + \epsilon \right) + P + b_c b^c \right] 
           \left( n^a u_a \right)^2 
          - \left[ P + {1\over2}\, b_c b^c \right] 
          - \left( n_a b^a \right)^2, \\ 
\fl S_b  &=&  - \left[ \rho_0 \left( 1 + \epsilon \right) + P 
         + b_c b^c \right] 
            \left( n^a u_a \right) \left( \perp\! u \right)_b 
         + \left( n_a b^a \right) \left( \perp\! b \right)_b, \\ 
\fl \left( \perp\! T \right)_{cd} &=&  
   \left[ \rho_0 \left( 1 + \epsilon \right) +\! P\! + b_c b^c  
   \right] 
   \left( \perp\! u \right)_c \left( \perp\! u \right)_d 
 + \left[ P + {1\over2} \, b_c b^c \right] h_{cd} 
 - \left( \perp\! b \right)_c \left( \perp\! b \right)_d,
\end{eqnarray}
where we have defined
\begin{equation}
W  \equiv  - n^a u_a , \qquad
v^a  \equiv  { 1 \over W} \, \left( \perp\! u \right)^a,
\end{equation}
and $\left( \perp\! X \right)^a \equiv h^a{}_b X^b$ denotes a projection.
Note that $W$ is the Lorentz factor between the fluid frame and
the fiducial observers moving orthogonally to the spatial hypersurfaces.
In addition, $v^a$ is the (purely spatial) coordinate velocity of the fluid.
The matter equations are projected along and orthogonal to $n^a$, and
expressed in terms of the conserved variables
\begin{eqnarray}
\fl 0  =  - n^a \partial_a E + K E  
  - {1 \over \alpha^2} D_a \left( \alpha^2 S^a \right) 
  + \left( \perp\! T \right)^{ab} \, K_{ab}, \\
\fl 0  =  h_{bc} \left[ - n^a \partial_a S^b + K S^b + 2 S^a K_a{}^b 
  - {1\over\alpha} S^a \partial_a \beta^b 
  - {1\over \alpha} D_a \left( \alpha \left( \perp\! T \right)^{ab} \right) 
  - {\partial^b \alpha \over\alpha} \, E \right], \\
\fl 0  =  D_a \left( * F^{ab} n_b \right), \\ 
\fl 0  =  h_{bc} \left[ - n^a \partial_a \left( *F^{de} n_d h^b{}_e \right) 
  + *F^{db} n_d K 
  + {1\over\alpha} D_a \left( \alpha \, (\perp\! *F)^{ab} \right) 
  - {1\over\alpha} \, {*F}^{da} n_d \partial_a \beta^b \right], \\ 
\fl 0 = {1\over \alpha} n^a \partial_a \left( \alpha D \right) + {1\over \alpha} \, D_a \left(\alpha D v^a \right) - K W ,
\end{eqnarray}
where $\alpha$ and $\beta^b$ are the ADM (3+1) lapse and shift, $K_{ab}$ is the 
extrinsic curvature, and $D_a$ is the covariant derivative compatible with 
$h_{ab}$.  These equations, in order, are the energy equation, the Euler 
equation, the no monopole constraint, the induction (or Faraday) equation 
and the baryon conservation equation.  

It is advantageous to use the 
standard magnetic field  as the evolution variable, rather than the magnetic
four vector $b^a$.  This amounts to working
in the frame of the fiducial observers moving along $n^a$ instead of in the 
fluid frame.  The electric
and magnetic fields in this frame are then 
\begin{equation}
E_a  =  h_a{}^b F_{bc} n^c, \qquad
B_a  =  {1\over2} \, \epsilon_{abc} F^{bc}.
\end{equation}
where $\epsilon_{abc} \equiv n^d \epsilon_{dabc}$.  The ideal MHD approximation then 
becomes a relation giving the electric field 
in terms of the magnetic field in the frame of the orthogonally moving 
observers:   
\begin{equation}
E_a = { 1 \over n_d u^d } \, \epsilon_{abc} u^b B^c.
\end{equation}

In practice, two modifications are made to the MHD equations in order to 
solve them.  First, we evolve the quantity $\tau = E - D$ instead of 
$E$ alone.  This is often done to have an energy quantity that reduces to the 
Newtonian value in the nonrelativistic limit.
Secondly, the source term in the induction equation can be eliminated
by combining that equation with the no-monopole constraint.
The final form for our matter equations thus becomes 
\begin{eqnarray}
\fl\partial_t \left( \sqrt{h} \, \tau \right) + \partial_i \left[ \sqrt{-g} \left( S^i
- {\beta^i \over \alpha} \, \tau - v^i D  \right) \right]  = \sqrt{-g} \, \left[ \left( \perp\! T \right)^{ab} \, K_{ab} - {1\over\alpha} \, S^a \partial_a \alpha \right],\\
\fl\partial_t \left( \sqrt{h} \, S_b \right) + \partial_i \left[ \sqrt{-g} \, \left( \left( \perp\! T \right)^i{}_b - {\beta^i \over \alpha} S_b \right) \right] \nonumber \\
\qquad\qquad = \sqrt{-g} \, \left[ \, {^{3}{\Gamma}}_{ab}^i \left( \perp\! T \right)^a{}_i + {1\over\alpha} S_a \partial_b \beta^a - {1\over\alpha} \partial_b \alpha \, E \right], \\
\fl- {1\over\sqrt{h}} \, \partial_i \left( \sqrt{h} \, B^i \right)  = 0,\\
\fl\partial_t \left( \sqrt{h} \, B^b \right) + \partial_i \left[ \sqrt{-g} \, \left( B^b \left( v^i - {\beta^i \over \alpha} \right) - B^i \left( v^b - {\beta^b \over \alpha } \right) \right) \right]  = 0 , \\ 
\fl\partial_t \left( \sqrt{h} \, D \right) + \partial_i \left[ \sqrt{-g} \, D \left( v^i - {\beta^i \over \alpha} \right) \right] = 0. 
\end{eqnarray}

\subsection{Primitive and conserved variables}
\label{subsec:primvars}

The evolution equations give the time dependence of the conserved
variables, $\bbu = ( D, S_i, \tau, B_j )^{T}$, but they also 
depend on the primitive variables $ \bbw = (\rho_0, v_i, P, b_j )^{T}$.  
As discussed in this section, for relativistic fluids the transformation
from conserved to primitive variables is transcendental.  The ability
to solve for physical values of the primitive variables under a wide
variety of conditions is an important and challenging part of writing
a relativistic fluid code.

The conserved variables are
\begin{eqnarray}
D  =  W \rho_0, \\
S_b  =  \left( h + b_c b^c \right) W^2 \, v_b + \left( n_a b^a \right) \left( \perp\! b \right)_b, \\
\tau  =    \left( h + b_c b^c \right) W^2
           - P - {1\over2} b_c b^c - \left( n_a b^a \right)^2 - W \rho_0, \\ 
B^a = - W b^a - u^a \cdot \left(n^c b_c \right) ,  
\end{eqnarray}
where the fluid enthalpy is $h= \rho_0(1+\epsilon) + P$. 
To obtain the inverse transformation, we reduce the problem to the solution for 
the roots of a single nonlinear function.  The method is as follows.

We eliminate the magnetic four vector, $b^i$, from the above equation 
using 
\begin{equation}
b^a = - {1 \over W} \left[ B^a + u^a \cdot (\perp\! u)^b B_b  \right].
\end{equation}
On replacing this, we get
\begin{eqnarray}
D  &=&  W \rho_0,\\
S_i  &=&  \left( hW^2 + B^2 \right) v_i - \left( B^j v_j \right) \, B_i,\\
\tau  &=&  hW^2 + B^2 - P - {1\over2} \left[ \left( B^i v_i \right)^2 
          + {B^2 \over W^2} \right] - W \rho_0, 
\label{eq:conserved_vars}
\end{eqnarray}
where $B^2 \equiv B_i B^i$, $v^2 \equiv v_i v^i$, and the indices are 
raised and lowered by the spatial metric $h_{ij}$.  
The spatial norm of $v^i$ can be expressed in terms of the Lorentz factor
\begin{equation}
W^2 = { 1 \over 1 - v^i v_i}.
\end{equation}
Density and pressure, two primitive variables, can be expressed as 
\begin{eqnarray}
\rho_0 = D \, {1\over W} =  D \, \sqrt{1 - v^2},   \qquad 
P = \left( h - \rho_0 \right) \, {\Gamma - 1 \over \Gamma}.
\end{eqnarray}
Note that we assume in this section a $\Gamma$-law equation of state.

It now remains to find $v^i$ (or $W$) and $h$ from our knowledge of
$D, S_i, \tau$ and $B_i$.  We contract $B^i$ with $S_i$
\begin{equation}
S_i B^i = h W^2 \, ( B^i v_i ),
\end{equation}
and use this to eliminate $B^i v_i$ in the expressions above
for $\tau$ and $S_i$.  From $S^i S_i$ we derive the expression
\begin{eqnarray}
\fl&- (hW^2)^2 \, W^2 \, S_i S^i + (hW^2)^2 \left( hW^2 + B^2 \right)^2 
\left( W^2 - 1 \right)\nonumber\\
\fl&\qquad\qquad\qquad\qquad\qquad\qquad - W^2 (2hW^2 + B^2) \left( S^i B_i \right)^2 = 0.
\end{eqnarray}
This can be solved for $W^2$ in terms of conservative variables and the 
quantity $x\equiv hW^2$:
\begin{equation}
W^2 = \left[ 1 - { (2x + B^2) (B^j S_j)^2 + x^2 (S^j S_j) \over x^2 (x + B^2)^2
} \right]^{-1}.
\label{eq:wx}
\end{equation}
Finally, we substitute \eref{eq:wx} into the equation for $\tau$ (which
comes about on using our above expressions for the density and pressure):
\begin{equation}
\fl \left[  x \left( 1 - { \Gamma - 1 \over \Gamma} \, {1 \over W^2} \right)
          - D \left( 1 - {\Gamma - 1 \over \Gamma} \, {1 \over W} \right)
          - \tau
          + {1\over 2} B^2 \left( 1 + v^2 \right)
    \right] x^2
  = {1 \over 2} \left( B^j S_j \right)^2.
\label{eq:x}
\end{equation}
The full expression is thus a nonlinear function in $x$,
 the roots of which we must calculate.  Note that all the coefficients in 
this expression are conservative variables that on numerical integration 
of the evolution equations will be known at a given time level.  
Once $x$ is obtained by solving \eref{eq:x}, it is then straightforward to 
find $W^2$, $v^2$, $h$, $\rho_0$, $P$ and $b^a$.  
\Eref{eq:x} is solved for $x$ numerically using a combined
Newton--Raphson and bisection solver.  In practice, a floor is placed on
$\rho_0$ and $P$, and a typical value for the floor is $10^{-10}$.  The
code simply halts when the primitive variable solver fails, and we do
not interpolate values of the primitive variables from neighboring points.

%
%
\section{Numerical methods}
\label{sec:numerical}

This section describes the numerical methods used to integrate 
the MHD equations.  
The fluid equations are solved using the Convex ENO  (CENO)
method~\cite{LiuOsher,DelZanna:2002qr}.  This method
is based on point values (FD discretization) rather 
than cell averages (FV discretization), simplifying
communications between grids.  FD fluid methods are advantageous 
for multiple domain problems in general relativity when the Einstein 
equations are discretized with finite differences.
A general relativistic MHD code with AMR, for example, can be simplified
if the fluid and geometric variables are refined in the same manner.

A second advantage of the CENO scheme is that the extension to systems
of equations uses a component-wise decomposition, rather than one based
on characteristic fields.  This eliminates the need to calculate left and
right eigenvectors of a Jacobian matrix.  Centered schemes are more
diffusive than those based on characteristic decompositions, but are
easier to implement numerically and more efficient.  Recent results show
that these methods work well with relativistic 
fluids~\cite{Lucas-Serrano:2004aq,Shibata:2005jv}.  Furthermore, 
the spectral decomposition of the Jacobian matrix for MHD is complicated
by the existence of various degeneracies, and centered schemes have been
widely used in relativistic 
MHD~\cite{Gammie:2003rj,DelZanna:2002rv,Duez:2005sf,Shibata:2005gp}.

In a free evolution of the MHD equations, the divergence of the magnetic
field can become very large.  Constrained transport, a discretization
technique for the magnetic field equations, is sometimes used to enforce 
the $\nabla\cdot{\bf B}=0$ constraint for the MHD equations.  
As we must interpolate data between arbitrary grids, we investigate two 
alternative 
methods for controlling this error.  The first method uses  an additional
hyperbolic field for divergence cleaning, and the second is an elliptic 
projection method.  Some comparisons
are made using both techniques for relativistic fluids.

Finally, we solve the equations on overlapping grids to facilitate the use
of uniform grids in complex geometries.  While the tests presented here are 
done in flat space, we use overlapping grids that mimic those used for 
excising black holes.

\subsection{Overlapping Grids}
\label{subsec:grids}

\begin{figure}
\begin{center}
\includegraphics*[height=6cm]{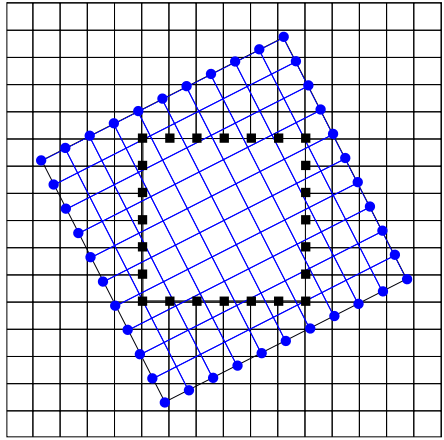}
\caption{This figure illustrates the overlapping grid structure used
in all of the numerical tests presented in this paper.  A region about
the origin is excised from the base grid, ${\cal G}_1$.  A second grid,
${\cal G}_2$, covers the excision region and is rotated with respect
to ${\cal G}_1$.  Data for the excision boundary of ${\cal G}_1$
are interpolated from ${\cal G}_2$, and outer boundary data for ${\cal G}_2$
are interpolated from ${\cal G}_1$.  Conventional out-flow boundary
conditions are applied to the outer boundaries of ${\cal G}_1$.
        }
\label{Fig:grids}
\end{center}
\end{figure} 

In many problems it is necessary or advantageous to choose coordinate
systems adapted to the boundaries.  Except for highly symmetric systems,
it is often difficult to choose a single coordinate chart appropriate for the
entire problem. 

A typical case of interest is a spacetime containing a black hole.
When black hole excision is used to remove singularities from the 
computational domain, adapting the boundary to the geometry of the
event horizon yields the maximal excision volume.
Coordinates adapted to the
event horizon or one black hole, however, are usually not appropriate 
for large domains or for binary black holes.   Thus multiple domain
numerical techniques are appropriate in numerical relativity
using both touching grids~\cite{Kidder:2000yq,Lehner:2005bz} 
and overlapping grids~\cite{Thornburg:2000cb,Thornburg:2003sf,%
Calabrese:2003vy,Calabrese:2004gs,Thornburg:2004dv}.  We concentrate
here on overlapping grids.

To test computational techniques for black hole spacetimes, we
perform all of the standard flat space tests in this paper on
overlapping grids with excision.  \Fref{Fig:grids} shows the basic
grid configuration used in these tests.  The grid ${\cal G}_1$ is
a base grid from which a square region is excised about the origin,
imitating the excision used around a black hole.  A second grid,
${\cal G}_2$, is placed over the excision region to give the usual simply
connected domain required for the flat space tests.  ${\cal G}_2$
is rotated by an arbitrary angle $\theta$ with respect to ${\cal
G}_1$.  We choose coordinates $(x,y)$ for ${\cal G}_1$ and coordinates
$(\xi,\eta)$ for ${\cal G}_2$, and the coordinates are related by a
simple rotation
\begin{equation}
\xi  = x\cos\theta + y\sin\theta, \qquad
\eta = -x\sin\theta + y\cos\theta.
\label{eq:grid_transform}
\end{equation}

During the evolution boundary data at the inner boundary of  ${\cal
G}_1$ are obtained by interpolating the solution from  ${\cal G}_2$.
Outer boundary data for  ${\cal G}_2$ are similarly obtained by
interpolating the solution on ${\cal G}_1$.  For clarity \fref{Fig:grids}
indicates these interpolated points in a single row at each boundary.
However, our third order evolution scheme requires a seven-point
stencil, and we actually interpolate a band of three points at each
inter-grid boundary.  Finally, interpolation zones between grids
are not allowed to overlap.

Conservative variables are interpolated at grid interfaces, although a
nonconservative interpolation scheme is used.  These
variables are typically smoother than the corresponding primitive variables,
and have smaller relative jumps near discontinuities.
The primitive variables are then recalculated from the interpolated 
variables.  Simple Lagrangian
interpolation can lead to oscillatory results near discontinuities,
frequently resulting in unphysical states in relativistic fluid dynamics.
(This effect may be less pronounced when using structured grids, such
for Berger-Oliger AMR.)
Therefore we interpolate with WENO interpolation~\cite{SebastianShu},
which is designed for use with discontinuous functions.
WENO interpolation is summarized in \ref{app:weno_interp}.

As mentioned briefly above, our FD fluid scheme on overlapping grids 
with WENO interpolation is
not conservative.  Conservative systems are often thought to be
necessary for obtaining the correct weak solutions to the fluid equations.
The effect of non-conservative boundary interpolation on such systems
has been examined by Tang and Zhou~\cite{TangZhou} and Sebastian and
Shu~\cite{SebastianShu}.  Tang and Zhou found that 
convergent weak solutions can be obtained using nonconservative 
interpolation at grid interfaces.  Sebastian and Shu found conservation
errors of second order at grid interfaces for smooth solutions and 
first order errors for solutions with discontinuities at the interface.
The numerical results presented in this paper are a direct demonstration
that we are able to
obtain the correct weak solutions when overlapping grids are used.

\subsection{CENO}
\label{subsec:ceno}

High-Resolution Shock-Capturing (HRSC) methods are designed to solve 
hyperbolic conservation laws of the form
\begin{equation}
\partial_t\bbu  + \partial_k\bbf\,^k(\bbu) = \bbs(\bbu),\qquad
\bbu(0,x^i) = \bbu_0(x^i),
\end{equation}
where $\bbu$  is a state vector, ${\bbf}\,^k$ are flux functions, and 
$\bbs$ contains source terms.  The semi-discrete discretization in one 
dimension is
\begin{equation}
\frac{d \bbu_i}{dt} 
 = - \frac{\skew 8\hat\bbf_{i+1/2} - \skew 8\hat\bbf_{i-1/2}}{\triangle x}
   + \bbs(u^i),
\label{eq:sd_eq}
\end{equation}
where $\skew 8\hat\bbf$ is a consistent numerical flux.  The accuracy of the
scheme depends, among other things, on the estimation, or reconstruction,
of $\bbu_{i+1/2}$ from the quantities 
$\bbu_{i-k},\cdots,\bbu_i,\cdots, \bbu_{i+m}$, where $k$ and $m$ are
integers.  
TVD schemes are named for the Total Variation Diminishing condition
on the interpolating polynomial.  These schemes accurately capture
the dynamics of strong shocks without oscillations or Gibbs overshoot
effects at discontinuities.  The TVD condition, however, can also
be overly restrictive, reducing the order of accuracy even at smooth
extrema.

The Essentially Non-Oscillatory (ENO) philosophy is that the 
interpolation stencil is chosen based on the local smoothness of
the function.  All points are used for smooth functions, and thus accuracy
is not lost at smooth extrema. Near discontinuities the stencil is
locally adjusted to use points away from the discontinuity.
In relaxing the TVD condition, small oscillations may develop
near discontinuities, but they are of $\rmO(\triangle x^k)$, where $k$
is the order of accuracy.  The flexibility of the ENO philosophy
has resulted in many extensions, including both FV and FD formulations, 
the Weighted ENO (WENO) approach, and schemes 
formally of very high order, etc.  Shu reviews the different
ENO methods in~\cite{ShuReview}.

The CENO method of Liu and Osher~\cite{LiuOsher} uses a FD discretization
and Lax--Friedrichs flux splitting, eliminating the need for a 
characteristic decomposition.  The interpolating polynomial in this
scheme is produced by a convex combination of lower order interpolations.
Near discontinuities this scheme is designed to produce results similar
to TVD methods.  This method was modified by Del Zanna and Bucciantini
for relativistic fluids~\cite{DelZanna:2002qr}, and we follow their approach.

We considered three different central or central-upwind numerical fluxes:
(1) the Lax--Friedrichs (LF) flux, (2) the local Lax--Friedrichs flux (LLF) and
(3) the Harten--Lax--van Leer (HLL) flux.  We are primarily 
interested in highly relativistic systems, where the characteristic speeds 
approach the speed of light.  In this limit the
HLL and LLF fluxes reduce to the simple LF flux.  Indeed,
we have found very little difference between solutions calculated
with the LF flux and those calculated with the LLF and HLL fluxes.
Similarly, Del Zanna \etal also reported nearly identical results from
the LLF and HLL fluxes in this regime~\cite{DelZanna:2002rv}.
In this paper we use only the LF flux
\begin{equation}
{\bbf}\,^{\rm LF}_{i+1/2} = \frac{1}{2}\left[ \bbf(\bbu^L_{i+1/2}) + \bbf(\bbu^R_{i+1/2})
          - (\bbu^R_{i+1/2} - \bbu^L_{i+1/2})\right],
\end{equation}
where $\bbu^L_{i+1/2}$ and $\bbu^R_{i+1/2}$ are the reconstructed states
to the left and right of the interface at $x_{i+1/2}$, respectively.
The CENO reconstruction for $\bbu^L_{i+1/2}$ and $\bbu^R_{i+1/2}$ 
is described in \ref{app:ceno_details}.

In FD ENO fluxes calculated from the point-wise 
values, $\bbf_{i+1/2}$,  must be converted into consistent numerical 
fluxes, $\skew 8\hat \bbf_{i+1/2}$.
To order $\mbox{O}(\triangle x^k)$, the conversion from point valued fluxes
to conservative fluxes is given by~\cite{ShuOsherI,ShuOsherII}
\begin{equation}
\skew 8\hat \bbf_{i+1/2} 
    = \bbf_{i+1/2} + \sum_{j=1}^{(k-1)/2}a_{2j}(\triangle x)^{2j}
\left(\frac{\partial^{2j}\bbf}{\partial x^{2j}}\right)_{i+1/2},
\end{equation}
where $a_2 = -1/24$ and  $a_4 = 7/5760$.
For second order schemes the two fluxes are identical, 
$\skew 8\hat \bbf_{i+1/2} = \bbf_{i+1/2}$.  
A third order scheme, however, requires the correction
\begin{equation}
\skew 8\hat \bbf_{i+1/2}= \left( 1 - \frac{1}{24}{\cal D}^{(2)}\right)
         \bbf_{i+1/2},
\end{equation}
where ${\cal D}^{(2)}$ is a second-order non-oscillatory difference operator.
${\cal D}^{(2)} f_i$ is calculated, again, as a convex combination of the 
differences
\begin{equation}
{\cal D}^{(2)}f_i = \mbox{minmod}
   (\alpha^{-1}D^{(2)}_- f_i, \alpha^0D^{(2)}_0 f_i,\alpha^{1}D^{(2)}_+ f_i),
\end{equation}
where the minmod limiter is
\begin{equation}
\mbox{minmod}(a_1, a_2, \cdots) = 
\left\{
\begin{array}{ll}
\min\{a_k\} & \hbox{if all $a_k > 0$,}\\
\max\{a_k\} & \hbox{if all $a_k < 0$,}\\
0 & \hbox{otherwise,}
\end{array}\right.
\label{eq:minmod}
\end{equation}
and the one-sided (first-order) and centered (second-order) 
second derivative operators are
\begin{equation}
\fl D^{(2)}_+ f_i = f_{i+2} - 2f_{i+1} + f_{i},\quad
D^{(2)}_- f_i = f_{i}   - 2f_{i-1} + f_{i-2},\quad
D^{(2)}_0 f_i = f_{i+1} - 2f_i     + f_{i-1}.
\end{equation}
The constants $\alpha^k$ are weights that may be chosen to bias towards 
centered differencing.  Here we use $\alpha^{-1} = \alpha^1 = 1$ 
and $\alpha^0 = 0.7$.

\subsubsection{Boundary conditions}

The third order CENO scheme has a seven point stencil.  At physical 
boundaries three ghost zones are used, which are populated by simple
extrapolation.  For example, at a boundary $x=x_0$ we set
$
u^n_{k,j} = u^n_{3,j}
$
for $k=0, 1, 2$ and all $j$.
At inter-grid boundaries these ghost zones are set by interpolating 
from a covering grid using WENO interpolation, as described 
in \ref{app:weno_interp}.

\subsubsection{Time integration}

The semi-discrete equations \eref{eq:sd_eq} are integrated in time 
using third order Runge-Kutta.  Various versions of RK3 exist, and we 
use the optimal third-order scheme of Shu and Osher that
preserves the TVD condition~\cite{ShuOsherI}
\begin{eqnarray}
\bbu^{(1)} &=& \bbu^{n} + \triangle t L(\bbu^n),\nonumber\\
\bbu^{(2)} &=& \frac{3}{4}\bbu^{n} + \frac{1}{4}\bbu^{(1)} 
           + \frac{1}{4}\triangle t L(\bbu^{(1)}),\\
\bbu^{n+1} &=& \frac{1}{3}\bbu^{n} + \frac{2}{3}\bbu^{(2)} 
           + \frac{2}{3}\triangle t L(\bbu^{(2)}).\nonumber
\end{eqnarray}
%

\section{The $\nabla \cdot {\bf B} = 0$ constraint}
\label{sec:divB}

The magnetic field, $\bf B$, must satisfy both its evolution equation as well
as the solenoidal constraint:   
$\nabla \cdot {\bf B} = 0$.  Small numerical errors lead to violations
of this constraint that, experience has shown, can grow rapidly.   Left
unchecked, violations of this constraint produce unphysical behavior.
Various approaches can be used to enforce the solenoidal constraint,
and here we consider two that can be applied to domains with multiple
arbitrary grids and with a view to incorporating AMR:  hyperbolic divergence 
cleaning and an elliptic projection
of $\bf B$.  We do not consider constrained transport here because it
requires that neighboring grids align in a structured manner, precluding
its application to overlapping grids with arbitrary coordinates, 
resolutions and/or orientations.

\subsection{Hyperbolic divergence cleaning}
\label{subsec:hyper_div_clean}

Hyperbolic divergence cleaning is simple to implement numerically
and comes with very little computational cost. 
A new scalar function $\psi$ is added to the
system, which can be interpreted as a generalized Lagrange multiplier (GLM),
and coupled to the magnetic field equations.   The method can be
implemented in various ways, and we follow the GLM method of 
Dedner \etal\ for the classical MHD equations~\cite{Dedner:2002}.
See van Putten for another approach~\cite{vanPutten}.

We specialize to Cartesian coordinates in flat space, and
assume that $\psi$ satisfies a linear differential equation and
couples to the magnetic field evolution equation according to 
\begin{eqnarray}
\label{eq:newB}
\partial_t B^b + \partial_i\left( B^b v^i - B^iv^b\right) 
           + g^{bj}\partial_j \psi &=& 0,\\
{\cal D}\psi + \nabla\cdot {\bf B} &=& 0,
\label{eq:psi}
\end{eqnarray}
where $\cal D$ is a linear differential operator.  Various choices for 
$\cal D$ can be made, giving hyperbolic, parabolic, and elliptic methods 
for divergence cleaning.  We choose
\begin{equation}
{\cal D}(\psi) = \frac{1}{c_h^2}\partial_t \psi + \frac{1}{c_p^2}\psi ,
\end{equation}
which combines some elements of both hyperbolic and parabolic operators, i.e.,
both propagating and damping the error.
\Eref{eq:psi} then becomes
\begin{equation}
\partial_t\psi + c_h^2\nabla\cdot{\bf B} = -\frac{c_h^2}{c_p^2}\psi.
\end{equation}
Differentiating and combining \eref{eq:newB}--\eref{eq:psi} shows that
$\psi$ also satisfies the telegraph equation
\begin{equation}
\partial_{tt}\psi + \frac{c_h^2}{c_p^2}\partial_t\psi - c^2_h\nabla^2\psi = 0.
\end{equation}
Violations of the solenoidal constraint propagate with the speed $c_h$ 
and the coefficient $c_p$ affects the damping rate.

We have tested hyperbolic divergence cleaning using the cylindrical shock
and relativistic rotor problems described below.  In our tests we found that
hyperbolic divergence cleaning to be very effective at keeping 
$||\nabla\cdot{\bf B}||_2$ bounded during an evolution.  Some numerical tests
are presented in \sref{sec:div_clean_tests}.
We turn now to the elliptic projection method. 

\subsection{The projection method}
\label{subsec:projection}

Blackball and Barnes~\cite{Brackbill} first proposed an elliptic projection 
correction to the magnetic field such that it satisfies the constraint 
$\nabla\cdot{\bf B} = 0$.  The evolution equations are used to obtain a 
preliminary estimate for the magnetic field  at the advanced time, 
${\bf B}^\star$.  The corrected magnetic field at the advanced time, 
${\bf B}^{n+1}$, is then obtained by solving the system
\begin{equation}
\nabla^2 \psi = \nabla\cdot {\bf B}^\star ,\qquad
{\bf B}^{n+1} = {\bf B}^\star - \nabla \psi.
\label{eq:proj}
\end{equation}
T\'oth has shown for classical MHD that (1) $\nabla\phi$ is the minimal
correction to ${\bf B}^\star$ that can be made such that 
$\nabla\cdot{\bf B}^{n+1} = 0$, and (2) the projection method gives 
the correct weak solution~\cite{Toth}.  

The projection method can be implemented in different ways, e.g., the
constraint can be imposed in Fourier space or physical 
space~\cite{Balsara:2003ui}.   We
consider here only two different discretizations in physical space.
Comparisons of different projection implementations, as well as to
other techniques, such as constrained transport, are given 
by T\'oth~\cite{Toth}, 
and Balsara and Kim~\cite{Balsara:2003ui}.
The divergence is discretized using the centered discrete 
operator $D_0$, and the Laplacian can be discretized with either $D_0 D_0$
or $D_+ D_-$.
In one dimension the operators are
\begin{equation}
(D_{0} v)_i   = \frac{v_{i+1} - v_{i-1}}{2\triangle x},\qquad
(D_+ D_- v)_i   = \frac{v_{i+1} -2v_i + v_{i-1}}{\triangle x^2}.
\end{equation}
The $D_0D_0$ operator has a five point stencil in each direction, and
the corrected magnetic field exactly satisfies the discrete divergence 
condition calculated using $D_0$.  The $D_+ D_-$ operator has a 
three-point stencil in each direction, and the corrected magnetic field does
not exactly satisfy any discrete divergence operator.  
As noted above, whether or not the magnetic field exactly satisfies a discrete
divergence condition is something of a red herring: the continuum solution
in both cases is only known to the level of truncation error.

The projection method requires the solution of an elliptic equation
\eref{eq:proj}, which we solve iteratively after each complete Runge-Kutta
cycle using a conjugate gradient  or stabilized bi-conjugate gradient method.  
The error tolerance for the elliptic solver
is set about 10 times less than the smallest expected truncation 
error, $\rmO(h^3)$.
The solution of the elliptic equations is relatively efficient, requiring
about 20--30\% of the total run time for typical resolutions.
Homogeneous Dirichlet boundary conditions are given for $\psi$
on all physical boundaries.  At overlapping grid boundaries $\psi$ is
interpolated as with other variables.  

When large corrections to ${\bf B}^\star$ are made in the projection
process, problems can arise in reconstructing the corrected primitive
variables.  This may occur because both $S_a$ and $\tau$ are 
themselves functions of the magnetic field, c.f.{} \eref{eq:conserved_vars}. 
We compensate by also ``correcting'' these variables after projecting $\bf B$,
slightly modifying the conventional correction used in classical 
MHD~\cite{Toth}.
The projection algorithm can thus be summarized by
\begin{enumerate}
\item Solve the evolution equations for preliminary values at the advanced
 time $\bbu^\star$;
\item From $\bbu^\star$ calculate the primitive variables at the advanced
 time $\bbw^{n+1}(\bbu^\star)$;
\item Solve \eref{eq:proj} for $\psi$ and compute ${\bf B}^{n+1}$;
\item Recalculate $\bbu^{n+1}$ from $\bbw^{n+1}$ and ${\bf B}^{n+1}$.
\end{enumerate}
This method assumes that the primitive variables more
accurately reflect the correct solution in the projection process.
Again, this is because $\rho$, $v^a$, and $P$ are not functions of
the magnetic field, whereas the conserved variables $S^a$ and $\tau$ are
functions of ${\bf B}$.

%
%
\section{Test Problems}
\label{sec:tests}

This section presents numerical results of our CENO scheme on overlapping grids.
The first tests are a set of standard Riemann problem tests for 
relativistic MHD proposed by 
Komissarov~\cite{Komissarov:2002mp,Komissarov:1999}.  Although these tests are
inherently one dimensional problems, we run them on unaligned 
overlapping grids, making them effectively two dimensional problems.
This allows us to test our divergence cleaning and interpolation methods
on known solutions.
We then discuss two test problems that are naturally two dimensional,
the cylindrical shock and relativistic rotor problems described by
Del Zanna \etal~\cite{DelZanna:2002rv}.  Finally we present comparisons
of the different divergence cleaning methods.

\subsection{Riemann problem tests}
\label{subsec:riemann_1d}

\begin{figure}
\begin{center}
\hbox{\includegraphics*[height=6.6cm]{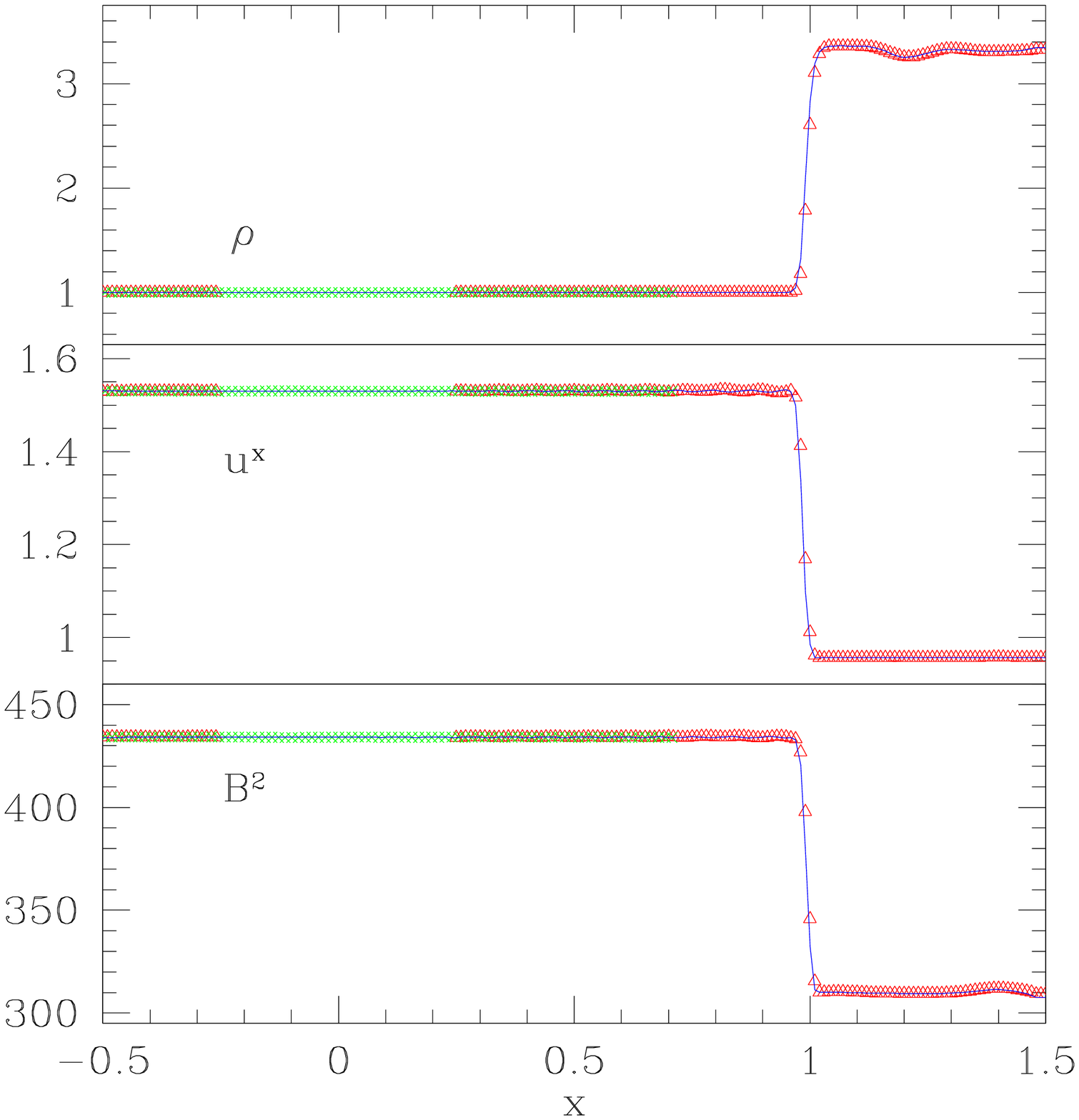}
\includegraphics*[height=6.6cm]{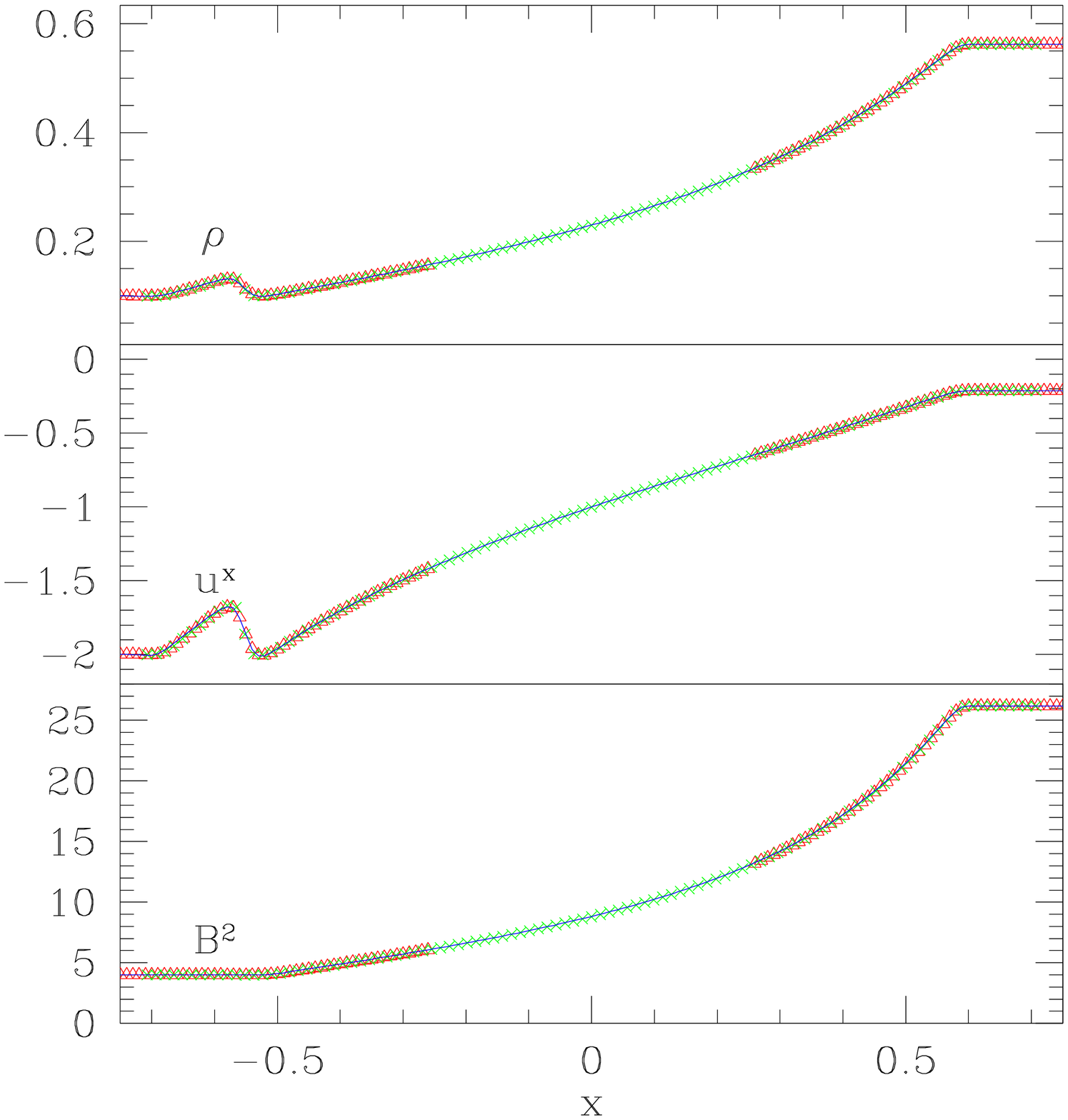}}
\hbox{\includegraphics*[height=6.6cm]{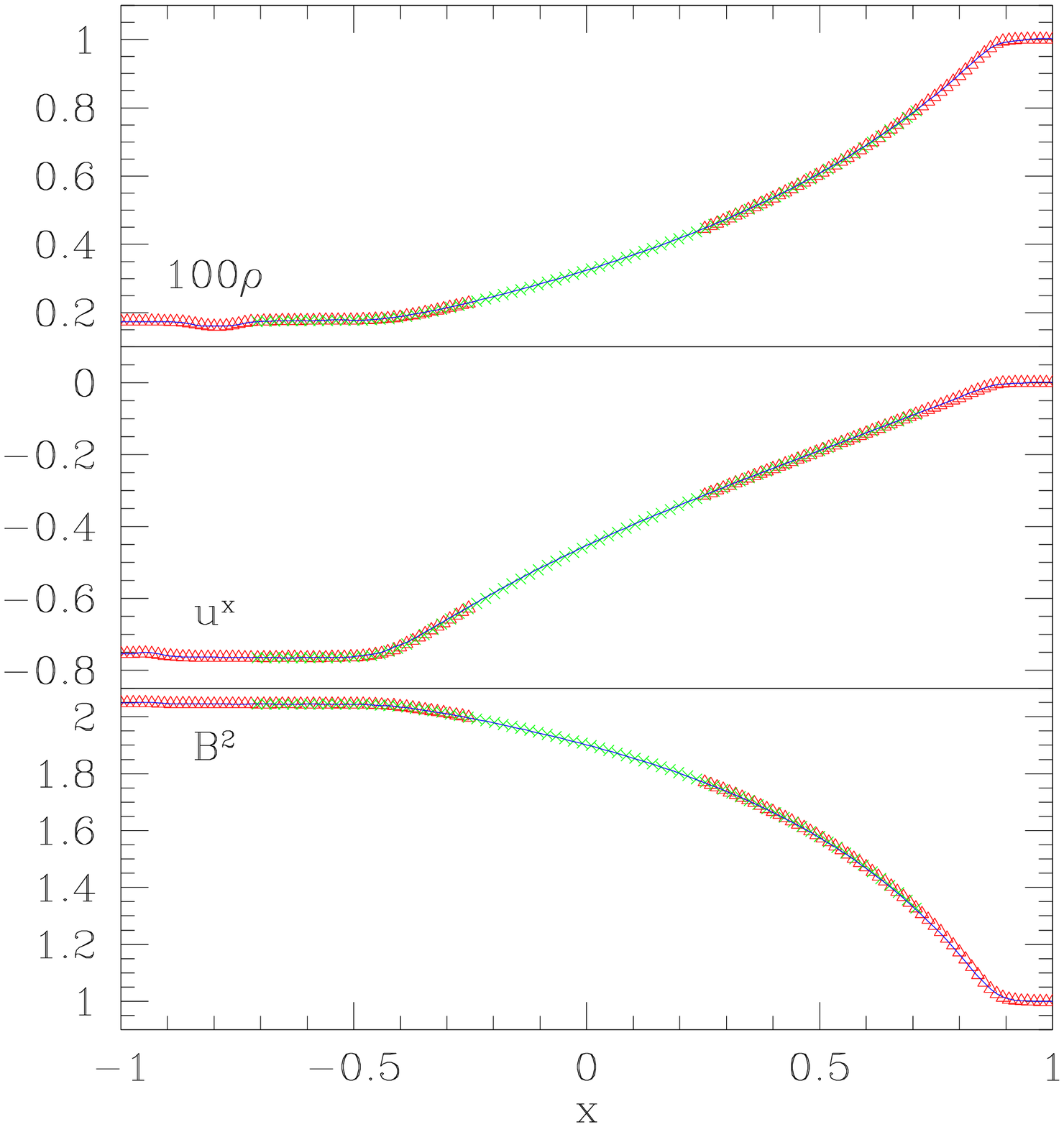}
\includegraphics*[height=6.6cm]{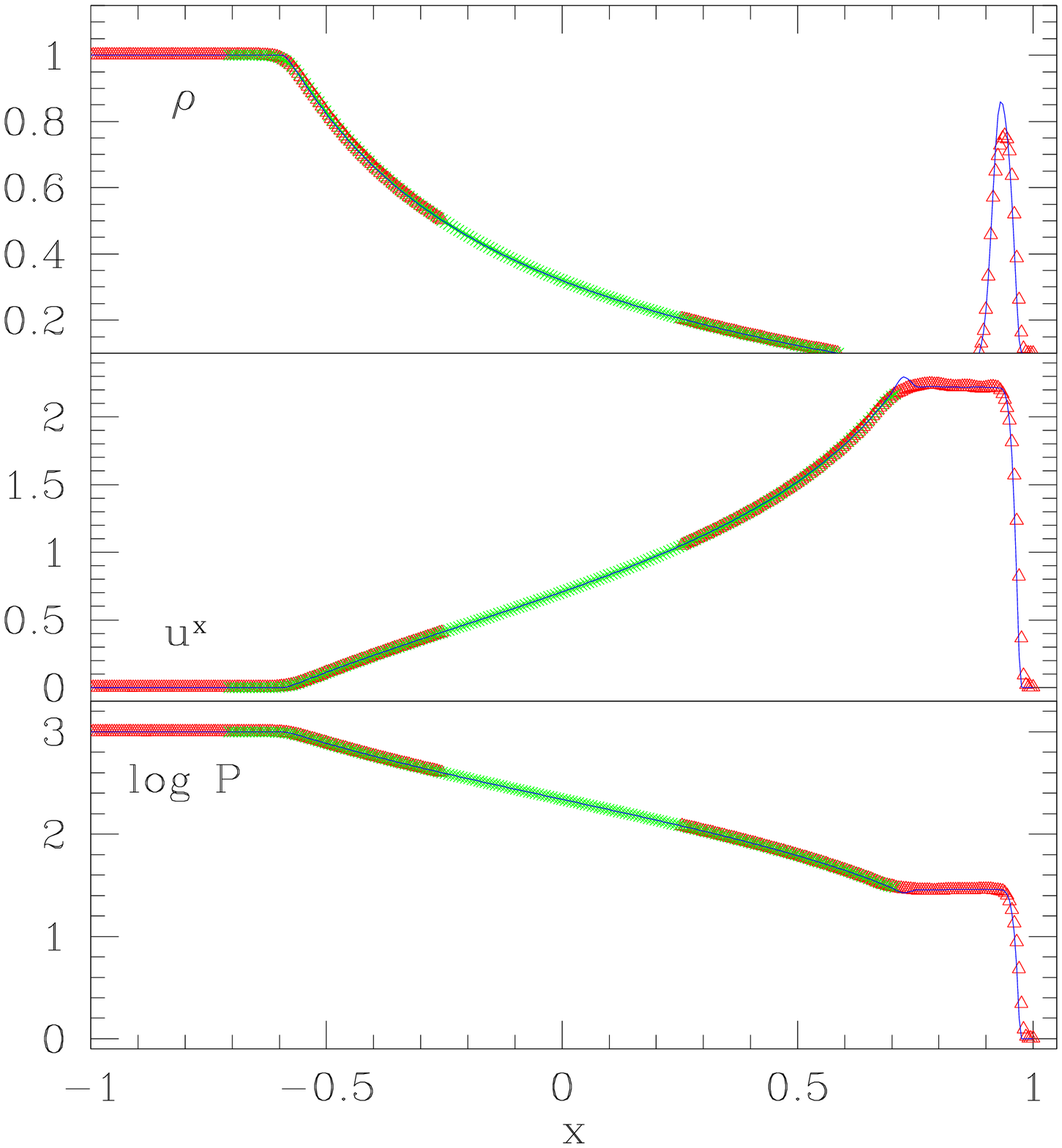}}
\caption{This figure shows four Komissarov test problems.  Solutions 
calculated on two-dimensional overlapping grids with elliptic divergence
cleaning (plotted along the line $y=0$) are compared with the one-dimensional 
solutions.  The solutions are nearly 
indistinguishable.
From left to right, top to bottom, these problems
are: (1) Slow Shock, (2) Switch-off Fast Rarefaction,
(3) Switch-on Slow Rarefaction, and (4) Compound Wave.
Triangles indicate the solution on the excised base grid,
${\cal G}_1$, and crosses indicate the solution on a grid covering
the excision region, ${\cal G}_2$,
rotated 45$^\circ$ with respect to the base grid.
Solid lines indicate the single-grid solution,
and not the exact solution.
        }
\label{Fig:fig_kst1}
\end{center}
\end{figure}

\begin{figure}
\begin{center}
\hbox{\includegraphics*[height=6.6cm]{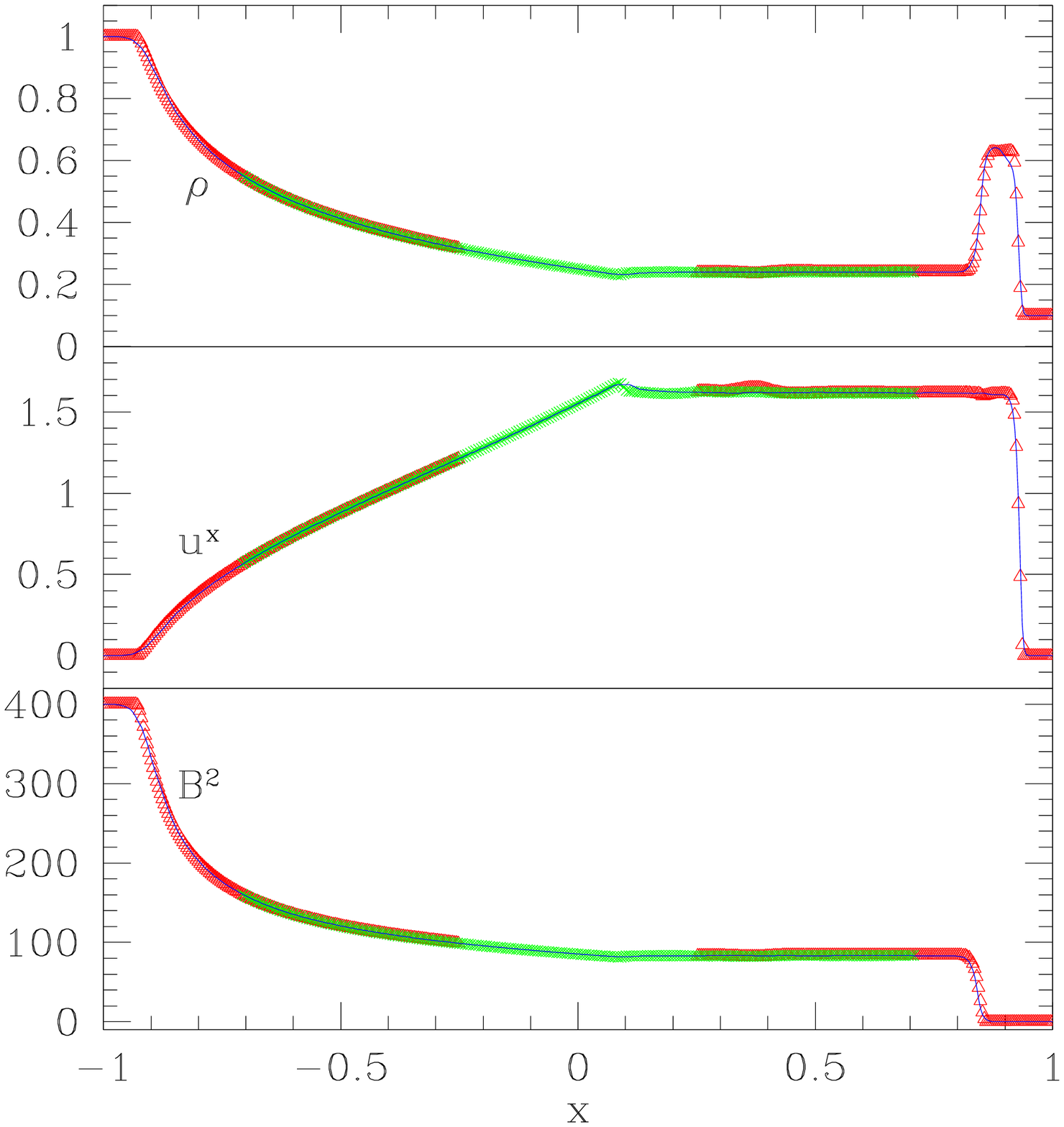}
\includegraphics*[height=6.6cm]{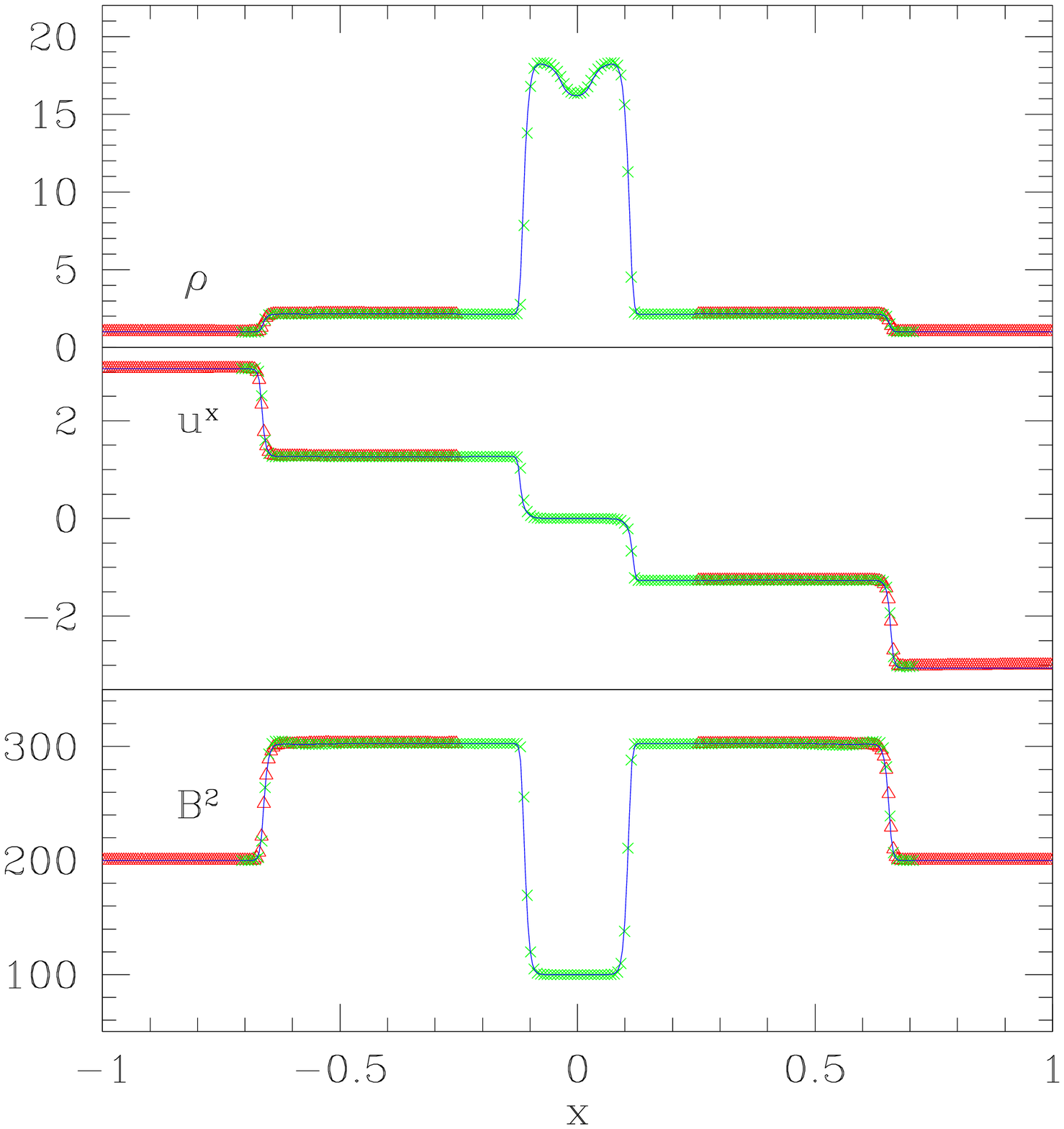}}
\caption{This figure shows two Komissarov test problems calculated on
overlapping grids.  The left frame shows Shock Tube \# 1 and the right frame 
is Shock Tube \# 2.  See the caption to \fref{Fig:fig_kst1} 
for further explanation.
        }
\label{Fig:fig_kst2}
\end{center}
\end{figure}

Solutions of the Riemann problem are used to test shock capturing numerical
methods, allowing one to verify that a method faithfully produces the 
fundamental shock, rarefaction and, for MHD,  Alfven waves.  
The analytic solution of the relativistic MHD Riemann problem is given
by Giacomazzo and Rezzolla~\cite{Giacomazzo:2005jy}.
Komissarov presented several Riemann problem 
tests for RMHD, which we have used to test our 
code~\cite{Komissarov:2002mp,Komissarov:1999}.  For comparison, results 
for Komissarov's tests have also been published
by other researchers~\cite{DelZanna:2002rv,Duez:2005sf,Shibata:2005gp}.
With the initial discontinuity aligned with a coordinate direction, we
are able to successfully reproduce all of Komissarov's test problems, 
though we have reduced the Courant number to $\lambda=0.4$ in all tests,
possibly because we are using the more diffusive Lax--Friedrichs flux.

When the discontinuity is rotated with respect to the coordinates, we are
able to run all Komissarov tests but two, the Fast Shock and
the Collision.  This problem occurs even when using a single computational
grid, and thus is not related to using overlapping grids.
In both cases we calculate unphysical values for the primitive variables,
and the code is immediately halted before the solution is completed.
We do not consider the Fast Shock problem here, and have modified the
Collision problem by reducing the initial velocities from
$\pm 0.981$ to $\pm 0.951$.

To provide the most comprehensive test of our algorithm, the Riemann problem
tests presented here are performed on unaligned overlapping grids with 
excision.  Excision is used here somewhat unconventionally, as the entire
computational domain is simply connected.  However, these Riemann problems
can be used to test the excision and divergence cleaning algorithms.  
For example, we test that 
unphysical effects do not arise as waves pass through grid interfaces,
and that the divergence cleaning methods do not adversely affect the solution.
These solutions are
then compared with those obtained from a single grid aligned to the initial
discontinuity, with no excision and where divergence cleaning is 
unnecessary.  Results from 
these tests are presented in Figures~\ref{Fig:fig_kst1} and \ref{Fig:fig_kst2}.

In all tests $\Gamma = 4/3$, the Courant factor is $\lambda=0.4$, and
elliptic divergence cleaning is used.
Second order reconstruction with the minmod limiter is used in all 
tests except the collision problem, where first order reconstruction is
more appropriate.  The initial discontinuity in the fluid data
is aligned with the coordinates $(x,y)$ of a base grid, ${\cal G}_1$.
The region $(x,y) \in [-\frac 1 4, \frac 1 4]$ is excised from ${\cal G}_1$,
and a second grid, ${\cal G}_2$, covers this excision region to form
the complete computational domain.  ${\cal G}_2$ has coordinates 
$(\xi,\eta)\in[-\frac 1 2, \frac 1 2]$, which are related to $(x,y)$
by a rotation of angle $\theta$ about the origin \eref{eq:grid_transform}.  
We choose $\theta=45^\circ$ for simplicity in plotting the results
of our runs.  In Figures \ref{Fig:fig_kst1} and \ref{Fig:fig_kst2} we plot 
data from the line $y=0$ for ${\cal G}_1$ and the diagonal elements 
of ${\cal G}_2$.

In all cases we see that the solutions calculated on overlapping grids
are nearly indistinguishable from those calculated on the single grid.
In particular, we do not see reflections from the grid boundaries.
Ill effects from elliptic divergence cleaning are also not observed
in comparison to the single grid runs, where divergence cleaning is not
used.

\subsection{Cylindrical blast wave}
\label{subsec:cylshock}

\begin{figure*}
\centerline{\vbox{
     \hbox{\includegraphics*[height=7.3cm]{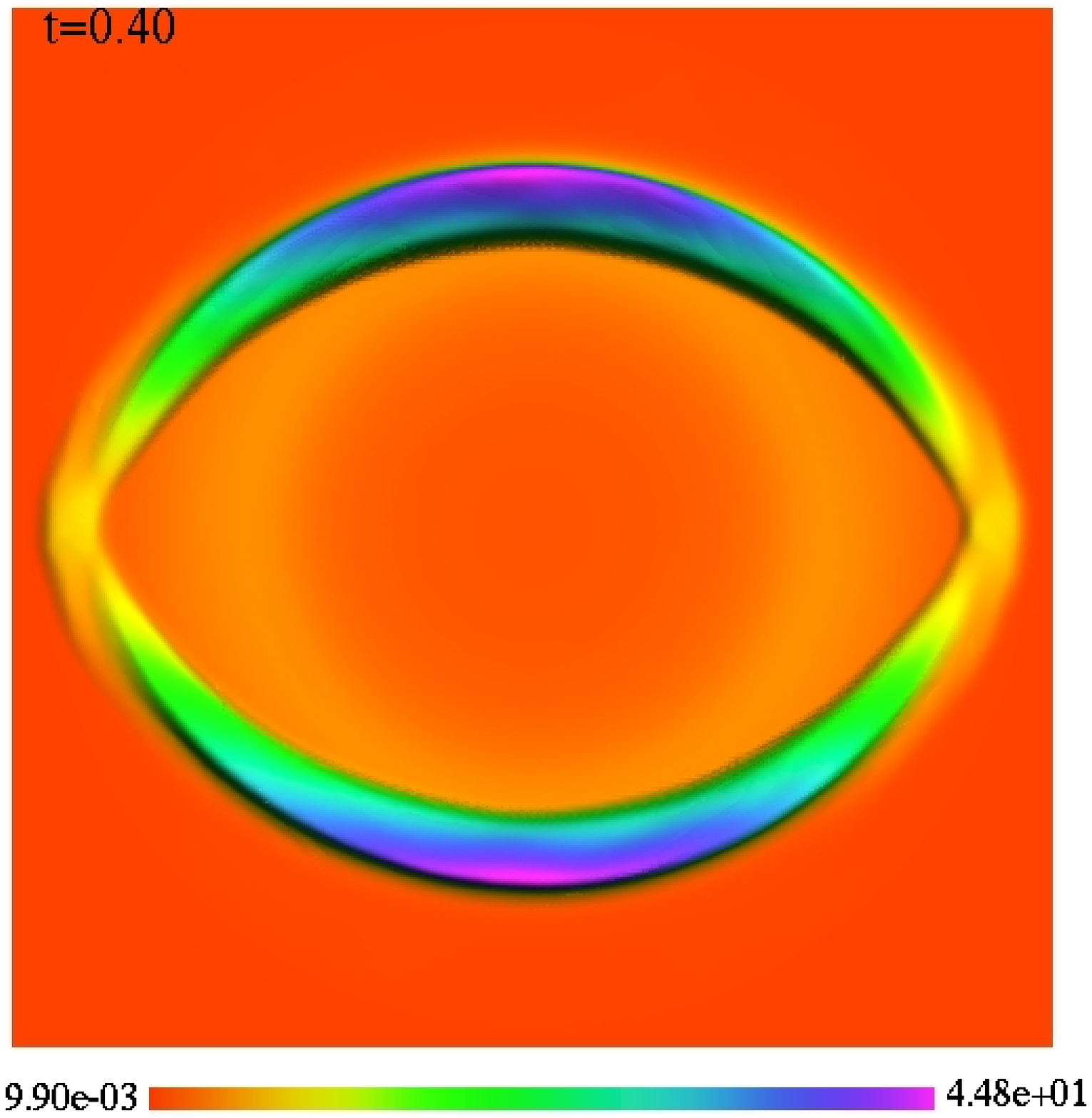}}
     \hbox{\includegraphics*[height=7.3cm]{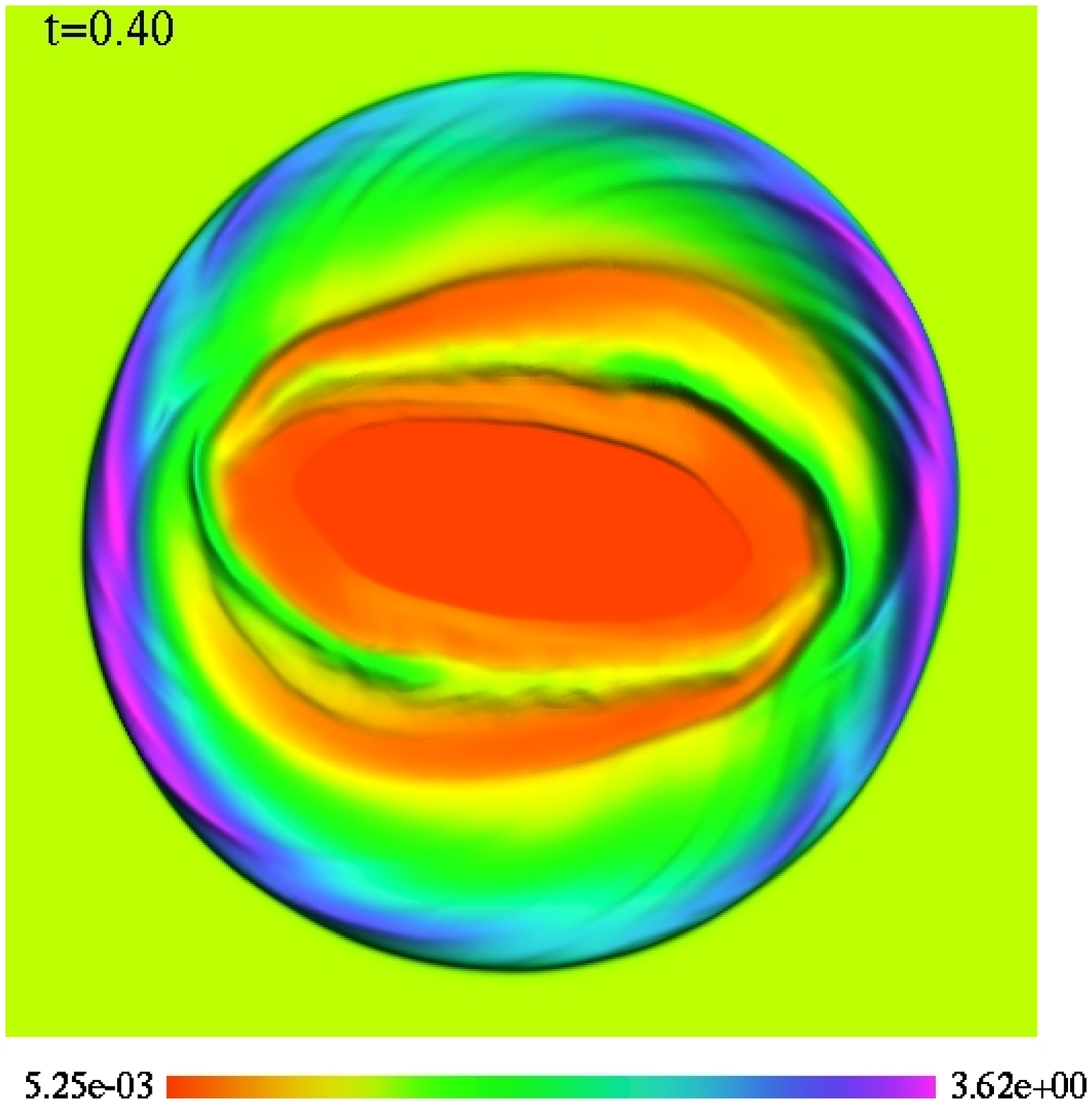}}
    }
    \vbox{
      \hbox{\includegraphics*[height=7.3cm]{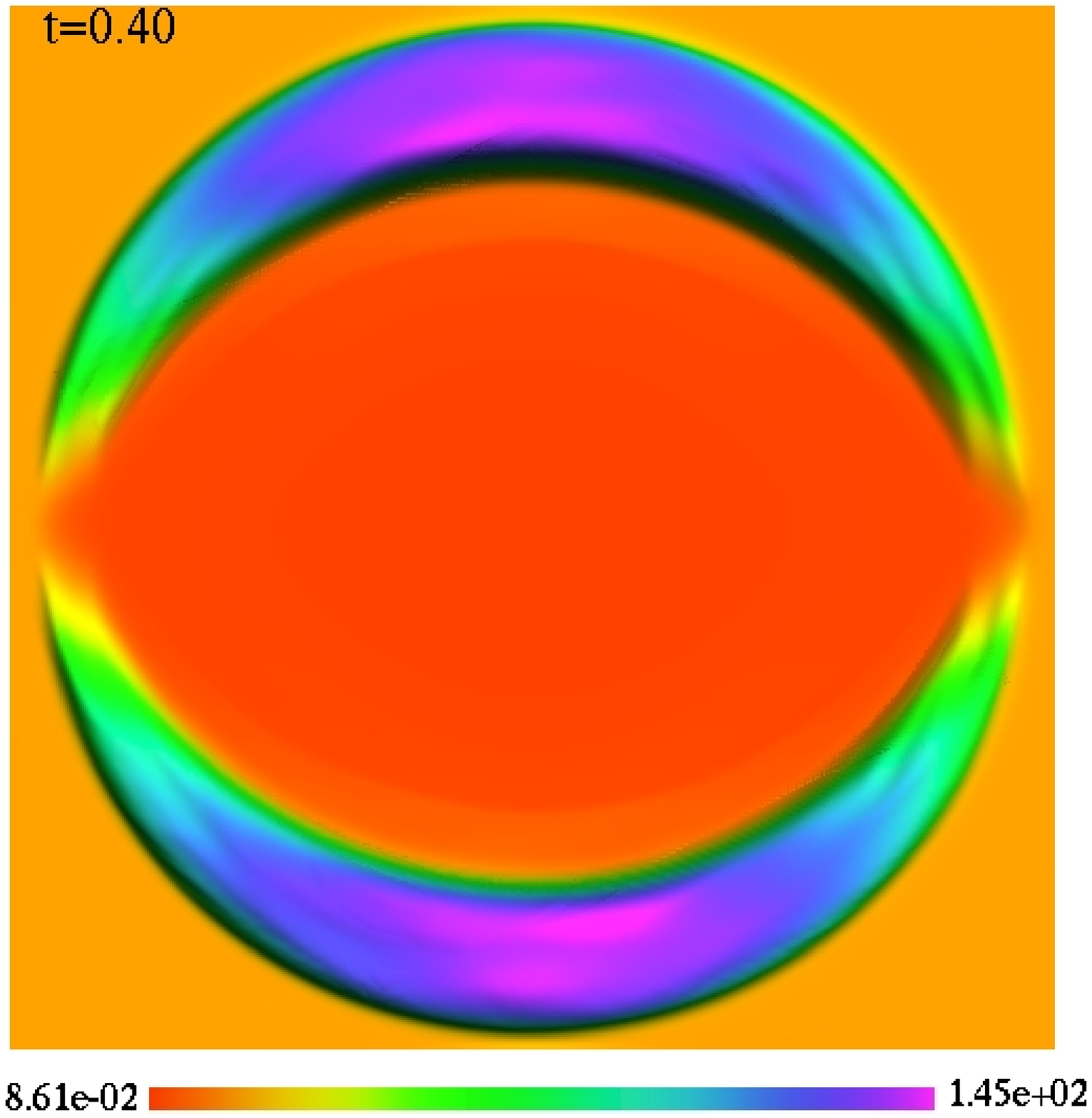}}
      \hbox{\includegraphics*[height=7.3cm]{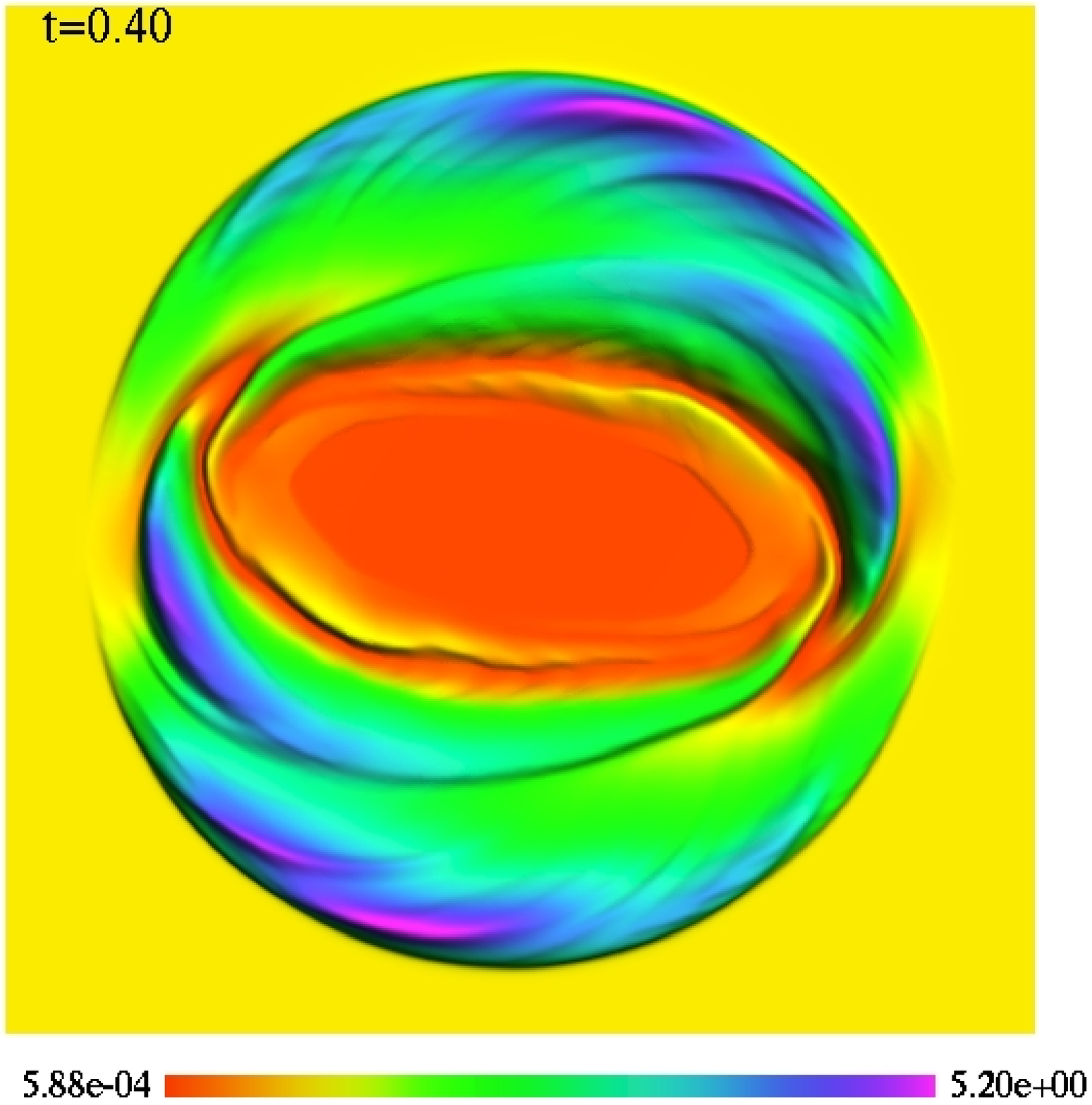}}
    }
 }
\caption{
This figure shows solutions of the cylindrical shock and relativistic
rotor test problems at $t=0.4$.
The top two frames show $P$ and $B^2$ for the cylindrical shock problem,
and the bottom two frames show the same variables for the relativistic 
rotor problem.  Details about the evolutions are given in the text.
}
\label{Fig:OGtests}
\end{figure*}

Del Zanna \etal presented this cylindrical shock
test problem for relativistic MHD~\cite{DelZanna:2002rv}.
The initial data consist of a uniform fluid background with
$\rho=1$, $P=0.01$, ${\bf v} = (0,0,0)$ and ${\bf B} = (4,0,0)$.
Inside a disk of radius $0.16$ centered at the origin, we set
$P=1000$.  The adiabatic index is $\Gamma=4/3$.  An analytic solution
is not available, but comparisons can be made with other published
results~\cite{DelZanna:2002rv,Shibata:2005gp}

The solution is calculated using two overlapping grids with an excised 
region in the base grid.  The base grid is uniform with $(x,y) \in [-1,1]$,
excluding the region $(x,y) \in [-0.2, 0.2]$.  The resolution is $h=0.008$.  
A second grid covers the excision region with
coordinates $(\xi,\eta)\in[-0.34, 0.34]$, rotated $50^\circ$ with respect 
to the base grid.

Figure~\ref{Fig:OGtests} shows $P$ and $B^2$ at $t=0.4$.  The solution
is calculated using second order reconstruction, with a Courant number of 
$\lambda=0.2$, and elliptic divergence cleaning.  The Courant number
is lower than that used by Del Zanna \etal, which may be a consequence
of a different numerical flux and using a grid, ${\cal G}_2$,
rotated with respect to
the initial magnetic field.  The pressure difference is initially very large,
leading to a strong out-going shock.  The initially circular shock becomes
distorted through interaction with the magnetic field, giving the elliptical
profile observed in the figure.  Comparing these results with those
of other researchers, no artificial grid effects are
observed in the solution, which could arise from using excision and 
overlapping grids.

\subsection{Relativistic rotor}
\label{subsec:rotor}

A second two-dimensional MHD test is the relativistic rotor, which evolves
an initially rigidly rotating fluid in the presence of a magnetic 
field~\cite{DelZanna:2002rv}.   Initial data consist of a constant 
background state
with $\rho=1$, $P=1$, ${\bf v} = (0,0,0)$ and ${\bf B}
= (1,0,0)$.   Inside a disk of radius $0.1$ at the center of the domain, 
the density is $\rho=10$, and the fluid is rigidly rotated with 
$\omega = 9.95$.  The linear velocity at the edge of the disk is $0.995$
and $W=10$.  Finally, the adiabatic index is $\Gamma =5/3$.

The solution is calculated using two overlapping grids.
The first grid has coordinates $(x,y) \in [-1/2,1/2]$, excluding the
region  $(x,y)\in[-0.13,0.13]$.  
A second grid covers the excision
region, and is rotated $27^\circ$ with respect to the first grid.
This grid has coordinates $(\xi,\eta)\in[-0.2, 0.2]$.  The resolution
on both grids is $h=0.0025$.  \Fref{Fig:OGtests} shows $P$ and $B^2$
at $t=0.4$.  The Courant factor is $\lambda = 0.2$, and second
order  reconstruction 
and elliptic divergence cleaning are used.  Again, these results appear
very similar to other published solutions, and no artificial grid effects
are observed.

\subsection{Divergence Cleaning}
\label{sec:div_clean_tests}

\begin{figure}
\begin{center}
\hbox{\includegraphics*[height=6.6cm]{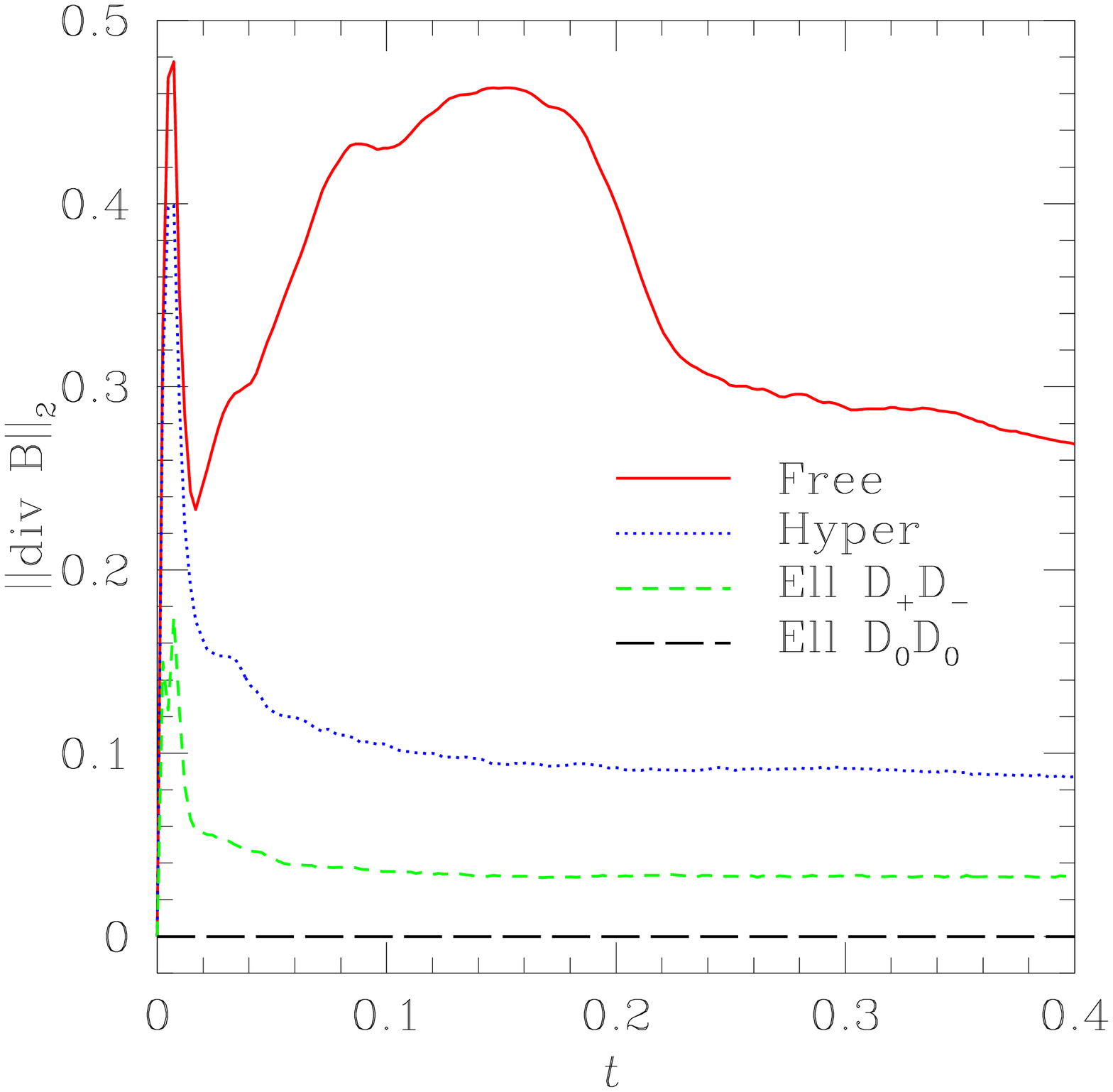}
\includegraphics*[height=6.6cm]{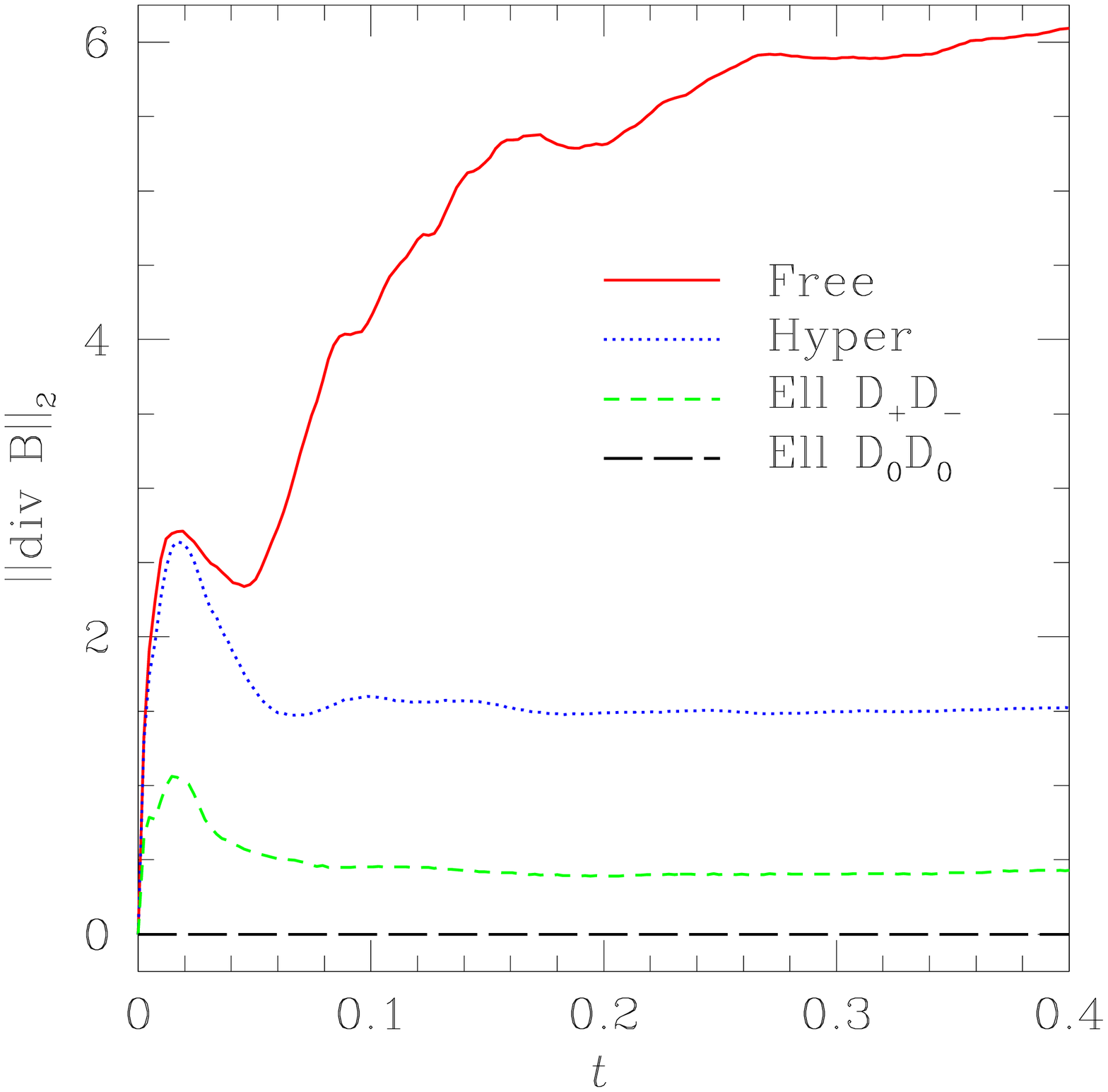}}
\caption{
This figure compares the hyperbolic and elliptic divergence cleaning methods
in two different cylindrical shock problems.  Each figure plots the L2 norm of 
$\nabla\cdot{\bf B}$ as a function of time for a free evolution, an evolution
using hyperbolic divergence cleaning, and elliptic divergence cleaning using
both $D_+D_-$ and $D_0 D_0$ operators.  Initial data for the left frame
have a central pressure $P_c=10$ and the central pressure for the right frame
is $P_c=1000$.  Unfortunately, these L2 norms are dominated by a few points
primarily near shocks, but this gives some indication of how violations of 
the constraint vary in time.
        }
\label{Fig:div_cleaning}
\end{center}
\end{figure}

\begin{figure}
\begin{center}
\hbox{\includegraphics*[height=6.6cm]{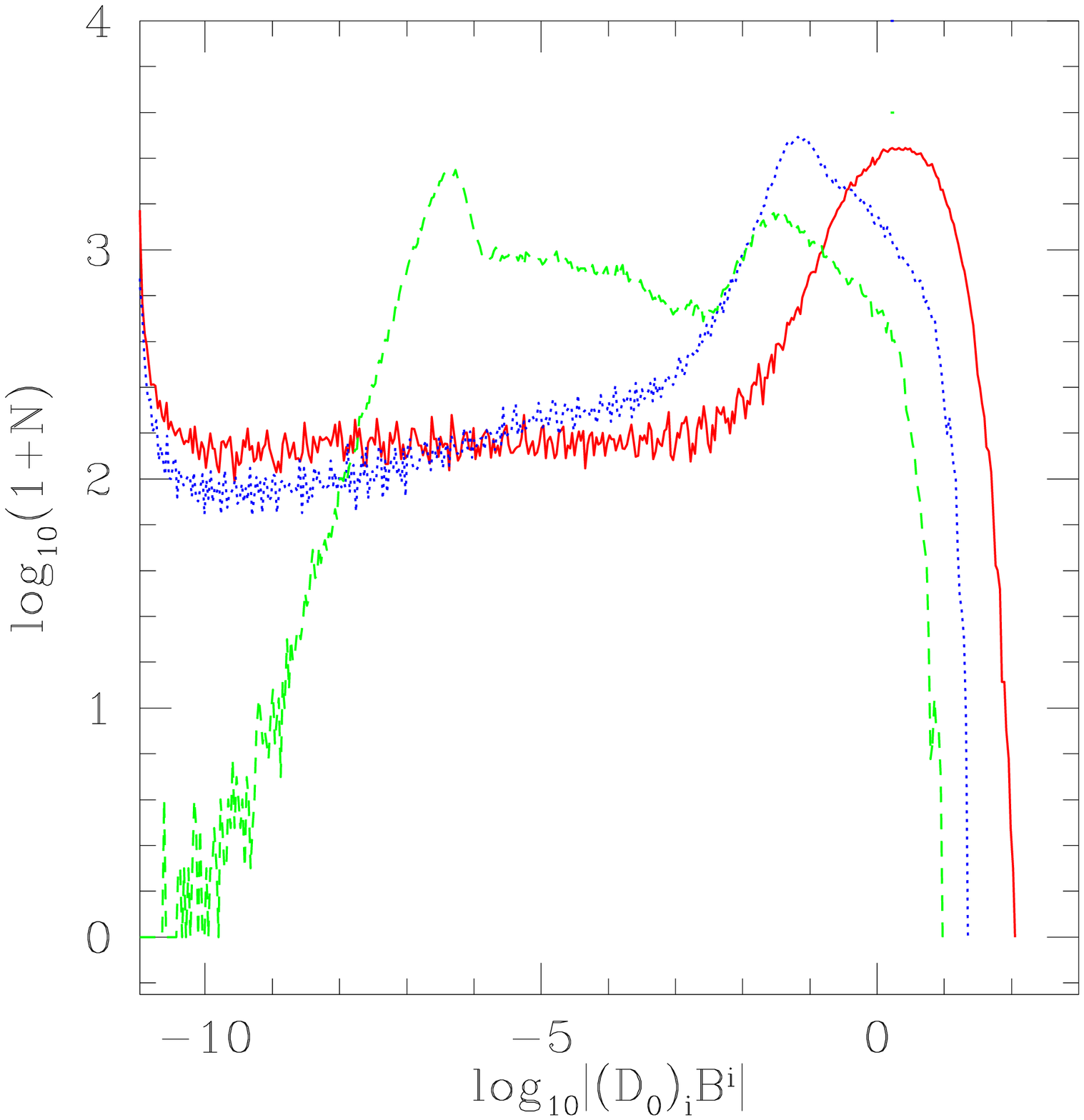}
\includegraphics*[height=6.6cm]{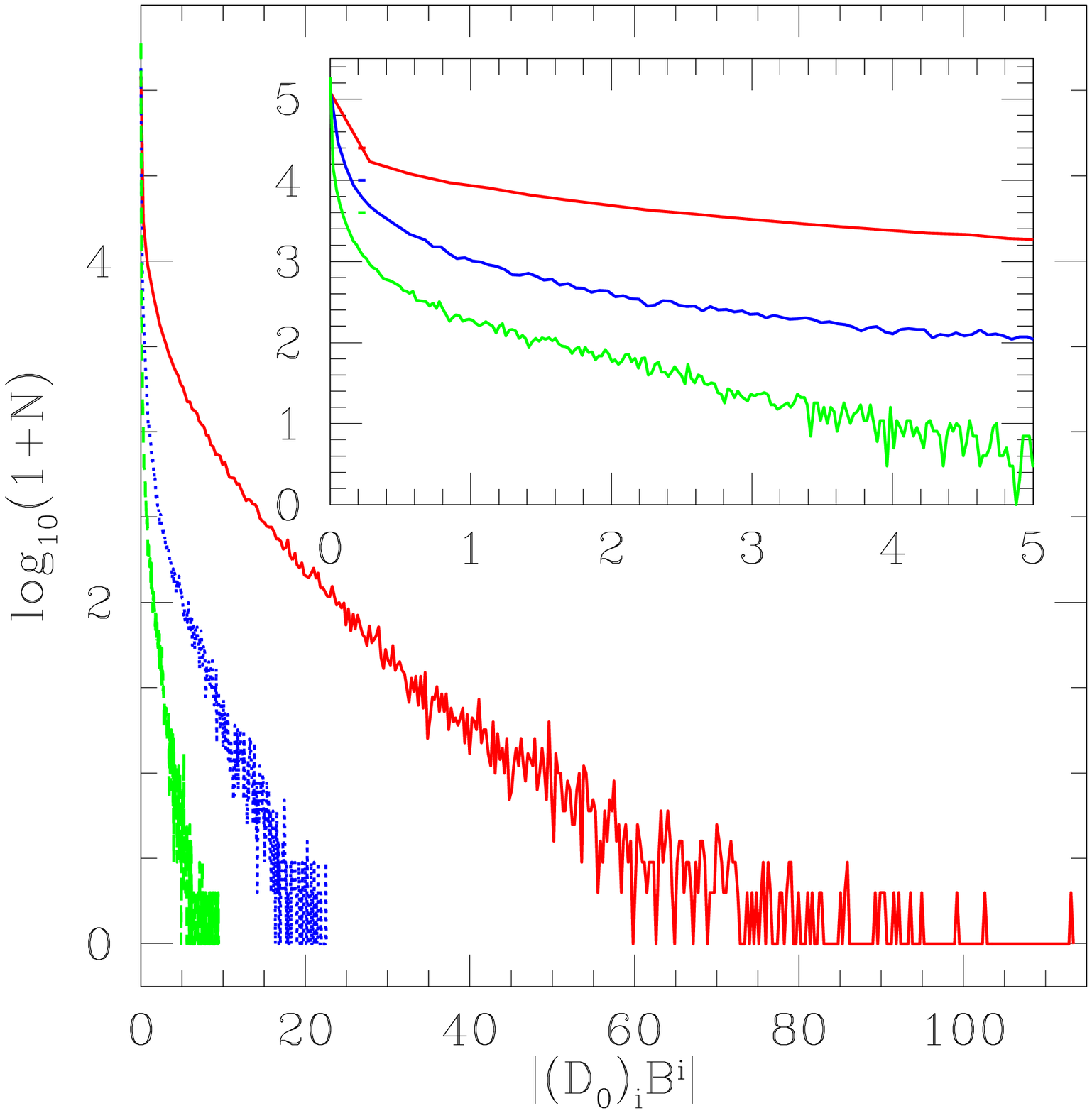}}
\caption{This figure shows the distribution of $|\nabla\cdot{\bf B}|$ 
at an instant of time, $t = 0.36$, for the cylindrical shock problem 
with $P_c=1000$.  On a given grid, the number of points, $N$, 
with $|\nabla\cdot{\bf B}|$ in a specified range are counted, 
and $\log_{10}(1+N)$ is plotted on the vertical axis. The total number
of points is $501^2$.  In the left frame 
the solid line corresponds to the
free evolution, the dotted line an evolution with hyperbolic divergence
cleaning, and the dashed line corresponds to elliptic divergence cleaning 
with the three-point stencil.  In the right frame, the lines are ordered
from top to bottom: free evolution, hyperbolic divergence cleaning, three-point
elliptic divergence cleaning.  Points with $|\nabla\cdot{\bf B}|<10^{-11}$
were excluded from the distribution on the left.
        }
\label{Fig:divB_spectrum}
\end{center}
\end{figure}

This section presents numerical results comparing the divergence cleaning
techniques discussed in \sref{sec:divB}.  We slightly modify the cylindrical
shock problem of \sref{subsec:cylshock}, and monitor $\nabla\cdot{\bf B}$
during the subsequent evolution for (1) a free evolution (no divergence 
cleaning), (2) an evolution with hyperbolic divergence cleaning, and
(3) elliptic divergence cleaning evolutions with $D_+D_-$ and $D_0D_0$
discrete divergence operators.

The cylindrical shock problem is modified by changing the background
pressure to $P=1$, and varying the central pressure, $P_c$.  In the examples
presented here, the central pressure is $P_c=10$ and $P_c=1000$. We
examined other values of $P_c$, but the results were essentially the same as
in these two cases.  (The differences arise only in the overall scale
of the constraint violation, not in the relative performance of each 
technique.)  The evolutions are performed on a single grid with
$501^2$ points with coordinates $(x,y)\in[-\frac 1 2, \frac 1 2]$.
During the evolutions $\nabla\cdot{\bf B}$ is calculated using the central
discrete operator $D_0$.  To emphasize the discrete nature of these
calculations, we write the numerically calculated divergences as $(D_0)_iB^i$.

Comparing the divergence cleaning techniques is more difficult than it
appears at first blush.  The first impulse is to simply plot L2 norms of 
$(D_0)_iB^i$ as a function of time, as shown in 
\fref{Fig:div_cleaning}.  Glancing quickly at this figure, one might conclude
that the elliptic divergence cleaning with $D_0D_0$ gives ideal
results, as $||(D_0)_iB^i||_2$ can be made as small as desired.
However, as discussed previously, ``exact'' satisfaction of the constant
can be a red herring, in that it does not imply that the continuum constraint
is ``exactly'' satisfied.  Using the same elliptic projection method with
a different discrete divergence operator, the $D_+D_-$ operator, gives a
non-zero value for $||(D_0)_iB^i||_2$ and a better indication of
error in the solution.

A second difficulty is that the largest constraint violations appear, not
surprisingly, near shocks.  Indeed, examination of the data show that a few
points near the shock completely dominate the values of  
$||(D_0)_iB^i||_2$.  This makes the comparison of L2 norms in 
\fref{Fig:div_cleaning} problematic, as they provide almost no information
about constraint violations in smooth parts of the solutions, where comparisons
between techniques may be more meaningful.  Moreover, since convergence
in the sense of Richardson extrapolation can not be defined for
discontinuous solutions, the norms $||(D_0)_iB^i||_2$ do not
become smaller with finer resolution, rather, the opposite occurs.
With finer resolution, the shock profile is sharpened, and derivatives of
the discontinuous variable approach the continuum derivative:
$d/dx[\theta(x-x')] = \delta(x-x')$.  Ideally one could remove points near
the shock from the comparisons, by either tracking the shocks or simply
removing points where $|(D_0)_iB^i|$ is judged to be too large.  We have no
facility for the former, and the latter strikes us as too arbitrary.
Since L2 norms of $(D_0)_iB^i$ give at best limited information, in
\fref{Fig:divB_spectrum} we plot the distribution of $|(D_0)_iB^i|$
at a single instance of time for the cylindrical shock problem with
$P_c=1000$.

%
%
\section{Conclusion}
\label{sec:conclusion}

The numerical scheme presented here is for solving the relativistic MHD
equations on multiple domains with overlapping grids.  While we
have not presented data from black hole spacetimes in this paper, it is the
target application that has influenced our design decisions.  First,
we choose to work with multiple domains since excision is most naturally 
implemented with smooth boundaries adapted to the event horizon's
geometry.  The flexibility of the overlapping grid approach allows one to 
easily use high resolution shock-capturing schemes on multiple domains.

Secondly, we choose an ENO method with a finite difference discretization
to simplify the transfer of information from one grid to another.  While
a conservative scheme for overlapping grids could be used~\cite{Liu2005}, 
working with point values instead of cell averages
simplifies the interpolation process for arbitrary grids.   
ENO finite difference schemes are also easily extended to higher dimensions 
and higher orders of accuracy.

Thirdly, we choose a central scheme to solve the RMHD equations.
Central schemes are very efficient HRSC methods, and ideal for combining
with AMR to resolve small features.

Fourth, we investigated the use of hyperbolic and elliptic divergence cleaning
to maintain the $\nabla\cdot{\bf B}=0$ constraint for MHD.  These techniques
are readily used on domains with arbitrary overlapping grids.  We found that 
hyperbolic divergence cleaning often gives acceptable results, especially
for moderate shocks.  Elliptic divergence cleaning is more robust, but
also more more computationally expensive.

Finally, it is natural to use a finite difference formulation of the
fluid equations when also solving the finite difference Einstein equations
with AMR.  When both the fluid and geometric variables are refined in the
same manner, the inconvenience of using staggered grids with AMR is
eliminated.

%
%
\ack

We would like to thank Matthew Anderson, Luis Lehner, Steven Liebling, 
Patrick Motl, Tanvir Rahman, Oscar Reula, and Joel Tohline for many 
interesting discussions during the course of this work.
This research was supported by the National Science Foundation under
grants PHY-0326378 and PHY-0502218 to Brigham Young University, and 
PHY-0244699 and PHY-0326311 to Louisiana State University.

%
%
\appendix

%
%
\section{CENO Reconstruction}
\label{app:ceno_details}

This appendix summarizes the CENO reconstruction scheme used by
Del Zanna, Bucciantini and Londrillo~\cite{DelZanna:2002qr,DelZanna:2002rv},
based on the original scheme of Liu and Osher~\cite{LiuOsher}.
The numerical fluxes are constructed dimension by dimension, thus the
basic algorithm is one dimensional.  Consider an uniform grid 
$x_i=i\triangle x$ with the function $v_i=v(x_i)$.
The standard one-sided and centered discrete differential operators are
\begin{equation}
(D_{\pm} v)_i = \pm \frac{v_{i\pm 1} - v_i}{\triangle x},\qquad
(D_{0} v)_i   = \frac{v_{i+1} - v_{i-1}}{2\triangle x}.
\end{equation}
To reconstruct $v_i$ on the interval $[x_{i-1/2},x_{i+1/2}]$ we first create
a linear TVD interpolating polynomial
\begin{equation}
v^{(1)}_i = v_i + v_i'(x-x_i),
\end{equation}
where $v_i'$ is the limited slope at $x_i$.  $v_i'$ is 
\begin{equation}
v'_i = \mbox{minmod}(D_- v_i, D_+ v_i).
\end{equation}
where the minmod limiter is defined in \eref{eq:minmod}.
Other TVD limiters can be considered, such as the monotonized central 
difference limiter, but we do not consider them here.
The first order reconstruction is $v(x)=v^{(1)}(x)$, which is equivalent
to the TVD reconstruction, and results in a second-order scheme.

Higher order reconstructions proceed hierarchically using the ENO philosophy 
of constructing multiple candidate polynomials, and then choosing the 
polynomial that is closest to the lower order polynomial.
For example
three candidate quadratic polynomials, $Q^k_i(x)$, $k=-1, 0, 1$,
for a second order reconstruction are
\begin{equation}
Q^k_{i}(x) = v_{i+k} + D_0 v_{i+k} (x - x_{i+k})
            + \frac{1}{2}D_+D_- v_{i+k} (x - x_{i+k})^2 .
\end{equation}
These second order polynomials are compared to the first order polynomial
at the point of interest, $x$, to calculate the weighted differences
\begin{equation}
d^k(x) = \alpha^k \left( Q^k_i(x) - v^{(1)}(x)\right).
\label{eq:eno_differences}
\end{equation}
The weights $\alpha^k$ are chosen to be $\alpha^{-1} = \alpha^1 = 1$ and
$\alpha^0 = 0.7$, biasing the interpolation towards the centered polynomial.
When all $d^k$ have the same sign, the second order reconstruction is
$v(x) = Q^\alpha_i(x)$, where $\alpha$ is the index corresponding to the
weighted difference with the smallest magnitude, 
$d^{\alpha}(x) = \mbox{min}(|d_k(x)|)$.
When $d_k(x)$ have differing signs, we revert to a first order reconstruction,
$v(x) = v^{(1)}(x)$.
This comparison to the first order reconstruction
yields results similar to TVD schemes near discontinuities, but gives an
higher order reconstruction for smooth functions.

%
%

\section{WENO interpolation}
\label{app:weno_interp}

\begin{figure}[t]
\begin{center}
\includegraphics*[height=2.2cm]{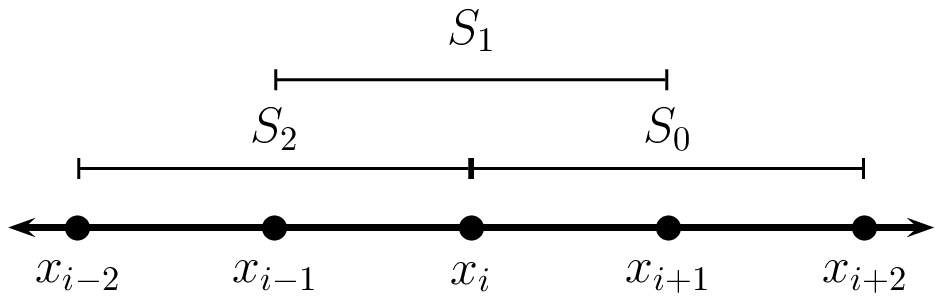}
\caption{A five point interpolation stencil is divided into three
substencils, $S_0$, $S_1$ and $S_2$, each with three points.  
Interpolants are calculated
using all three stencils, which are then combined in a weighted convex sum.
The nonlinear weights depend on the smoothness of the function.
        }
\label{Fig:weno_stencil}
\end{center}
\end{figure} 

Boundary data at interfaces between overlapping grids are obtained
by interpolation.  Following Sebastian and Shu~\cite{SebastianShu}, 
we have investigated
both Lagrangian and WENO interpolation at these grid interfaces.
Lagrangian interpolating polynomials work well for smooth functions.
Near discontinuities, however, they become oscillatory, which can lead
to unphysical states in relativistic MHD, i.e., physical primitive 
variables can not be obtained from the interpolated conservative 
variables.   WENO interpolation avoids these oscillations near
discontinuities by adjusting the interpolation stencil according
to the local smoothness of the data.  The WENO interpolant is 
constructed as a convex sum of lower order interpolations computed on 
substencils.  The contribution from each substencil
is weighted nonlinearly by measures of the function's local smoothness.
For smooth functions, WENO interpolation closely approximates Lagrangian
interpolation. A lower-order interpolant is calculated near discontinuities,
which is biased towards substencils where the function is smooth.
In this appendix we summarize 
the WENO interpolation algorithm~\cite{SebastianShu} that we use 
exclusively in our code.

Consider a discrete function $u_i$ defined at $(2k-1)$ points, $x_i$,
$i=0, \ldots , 2k-2$.  The Lagrangian interpolation polynomial $u_L(x)$ is
\begin{equation}
u_L(x) = \sum_{i=0}^{2k-2} c_i(x) u_i, \qquad
c_i(x) = \prod_{j=0}^{2k-2}\frac{x - x_j}{x_i - x_j}.
\end{equation}
To motivate WENO interpolation, we divide the $(2k-1)$ points
into $k$ substencils, see \fref{Fig:weno_stencil},
and write $u_L(x)$ as a sum of lower order
interpolating polynomials on each substencil.  Let the substencils be
$ S_r(i) = \{x_{i-r}, \ldots, x_{i-r+k-1} \}$ for  
$r = 0, \ldots , k-1$.
The Lagrangian interpolating polynomial on each substencil is
\begin{equation}
\fl u^{(r)}_L(x) = \sum_{j=0}^{k-1}u_{i-r+j}c_{rj}(x),\qquad
c_{rj}(x)  = \prod_{\ell=0, \ell \neq j}^{k-1} 
            \frac{x-x_{i-r+\ell}}{x_{i-r+j} - x_{i-r+\ell}}.
\end{equation}
We can expand $u_L(x)$ in terms of $u^{(r)}_L(x)$ as
\begin{equation}
u_L(x) = \sum_{r=0}^{k-1} d_r(x) u^{(r)}_L(x),
\label{eq:lpsum}
\end{equation}
where $d_r(x)$ are constants, or linear weights, that depend on $x$.
Consistency requires that
$
\sum_{r=0}^{k-1} d_r(x) = 1.
$

The Lagrangian interpolation polynomial $u_L(x)$ is written above as a sum
of lower-order polynomials in \eref{eq:lpsum}.  
WENO interpolation
generalizes this by creating a convex sum, where the weights are nonlinearly
dependent on the local smoothness of $u_i$.  The WENO interpolation
polynomial has the form
\begin{equation}
u_w(x) = \sum_{r=0}^{k-1} \omega_r(x) u^{(r)}_L(x)
\end{equation}
where $\omega_r(x)$ are the nonlinear weights, and for consistency
we require
$
\sum_{r=0}^{k-1} \omega_r(x) = 1.
$
Following the fundamental reconstruction procedure for WENO evolution
schemes, the weights are chosen to be
\begin{equation}
\omega_r(x) = \frac{\tilde \omega_r(x)}{\sum_{s=0}^{k-1} \tilde \omega_s(x)},
\qquad 
\tilde \omega_r(x) = \frac{d_r(x)}{(\varepsilon + \beta_r(x))^2},
\end{equation}
where $\varepsilon$ is a small number that we set as $\varepsilon=10^{-6}$.
The coefficients $d_r(x)$ are obtained from \eref{eq:lpsum}, and the 
functions $\beta_r(x)$ are smoothness indicators given by
\begin{equation}
\beta_r(x) = \sum_{\ell=1}^{k-1}\int_a^b \triangle x^{2\ell-1}
                      \left( \frac{d^\ell}{dx^\ell} u^{(r)}_L(x)
                      \right)^2\, \rmd x ,
\end{equation}
where the limits of integration are over different substencils.
All interpolations for the grids used in this paper are centered,
$x\in[x_{i-1/2},x_{i+1/2}]$. In this case the smoothness indicators
become~\cite{SebastianShu}
\begin{eqnarray}
\fl\beta_0 &=& (10 u_i^2 - 31 u_i u_{i+1} +25u_{i+1}^2 + 11 u_i u_{i+2}
         - 19u_{i+1}u_{i+2} + 4u_{i+2}^2)/3\\
\fl\beta_1 &=& ( 4 u_{i-1}^2 - 13 u_{i-1}i u_{i} + 13u_i^2 + 5u_{i-1}u_{i+1}
         -13u_i u_{i+1} + 4u_{i+1}^2)/3\\
\fl\beta_2 &=&  (4 u_{i-2}^2 - 19 u_{i-2}u_{i-1} + 25u_{i-1}^2 + 11u_{i-2}u_i
         - 31u_{i-1}u_i + 10u_i^2) / 3
\end{eqnarray}

\begin{figure}
\begin{center}
\includegraphics*[height=6cm]{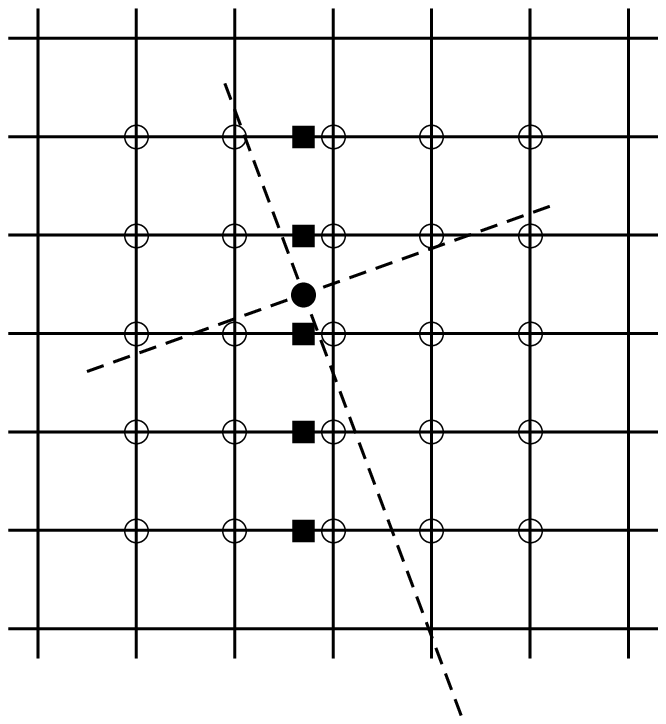}
\caption{Interpolations in two dimensions are calculated as a series
of one dimensional interpolations.  The dashed diagonal lines represent
coordinate lines for a grid that requires interpolated at their
intersection, indicated by the solid circle.  The interpolation data
are calculated from the grid with rectangular coordinate lines by
first interpolating horizontally, using the points indicated by open
circles, to obtain data at the points indicated by solid squares.  
The stencil indicated by solid squares
is then used to obtain interpolation data at the required point.
        }
\label{Fig:weno_interp}
\end{center}
\end{figure}

Since the interpolation coefficients $\omega_r(x)$ 
depend on the smoothness of the function, they must be calculated for each 
function individually,
whereas Lagrangian interpolation coefficients are only position dependent.
This makes WENO interpolation more expensive to implement when several
functions must be interpolated at a single point.
Finally,
Interpolations in two dimensions are calculated as a series of one dimensional
WENO interpolations, as shown in \fref{Fig:weno_interp}.

%
%


\section*{References}
\begin{thebibliography}{99}

\bibitem{Blandford:1977}
  Blandford R D and Znayek R L 1977
  MNRAS {\bf 179} 433

\bibitem{Balbus}
  Balbus S A and Hawley J F 1991
  {\it Astrophys.\ J.}  {\bf 376} 214

\bibitem{ADM}
Arnowitt R, Deser S and Misner C 1962 {\it Gravitation: An Introduction
    to Current Research,} ed L~Witten (New York: Wiley)

\bibitem{Sloan:1985}
  Sloan J and Smarr L L 1985
  {\it Numerical Astrophysics} ed J L J Centrella \etal
  (Boston: Jones and Bartlett) p 52

\bibitem{Evans:1988}
  Evans C R and Hawley J F 1988
  {\it Astrophys. J. } {\bf 332} 659

\bibitem{Koide:2000}
  Koide S, Meier D L, Shibata K and Kudoh T 2000
  {\it Astrophys.\ J.}  {\bf 536} 668

\bibitem{DeVilliers:2002ab}
  De Villiers J-P and Hawley J F 2002
  arXiv:astro-ph/0210518

\bibitem{Gammie:2003rj}
Gammie C F, McKinney J C and T\'oth G 2003
  {\it Astrophys. J.}  {\bf 589} 444

\bibitem{Baumgarte:2002vu}
  Baumgarte T W and Shapiro S L 2003
  {\it Astrophys.\ J.}  {\bf 585} 930

\bibitem{Komissarov:2004}
  Komissarov S S 2004 MNRAS {\bf 350} 1431

\bibitem{Anton:2005gi}
  Anton L Zanotti O Miralles J A, Marti J M, Ibanez J M, Font J A and Pons J A
  2005
  Numerical 3+1 general relativistic magnetohydrodynamics: a local
  characteristic approach
  {\it Preprint} astro-ph/0506063

\bibitem{Duez:2005sf}
Duez M D, Liu Y T, Shapiro S L and Stephens B C 2005
  Relativistic Magnetohydrodynamics In Dynamical Spacetimes: Numerical
  Methods And Tests
  {\it Preprint} astro-ph/0503420.

\bibitem{Shibata:2005gp}
  Shibata M and Sekiguchi Y I 2005
  \PR D {\bf 72} 044014

\bibitem{ShuReview}
  Shu C-W 1997 ICASE Report 97-65, NASA Langley Research Center

\bibitem{Scheel:2000}
  Scheel M 2000 Miniprogram on Colliding Black Holes: Mathematical Issues
    in Numerical Relativity, Institute for Theoretical Physics, University
    of California at Santa Barbara, January 10--28 2000.  Available online
    at http://online.kitp.ucsb.edu/online/numrel00

\bibitem{Lehner:2004}
 Lehner L, Neilsen D, Reula O and Tiglio M 2004 \CQG {bf 21} 5819

\bibitem{Calabrese:2003vy}
  Calabrese G and Neilsen D 2004
  \PR D {\bf 69} 044020

\bibitem{Kidder:2000yq}
  Kidder L E, Scheel M A, Teukolsky S A, Carlson E D and Cook G B 2000
  \PR D {\bf 62} 084032

\bibitem{Lehner:2005bz}
  Lehner L, Reula O and Tiglio M 2005
  \CQG {\bf 22} 5283

\bibitem{Starius}
Starius G 1980
  {\it Numer. Math.} {\bf 35} 241

\bibitem{Henshaw}
Chessire G and Henshaw W D 1990
  {\it J. Comput. Phys.} {\bf 90} 1

\bibitem{Calabrese:2004gs}
  Calabrese G and Neilsen D 2005
  \PR D {\bf 71} 124027

\bibitem{LehnerReula}
  Lehner L and Reula O 2005 Personal communication

\bibitem{Toth}
 T\'oth G 2000
 {\it J. Comput. Phys.} {\bf 161} 605

\bibitem{Balsara:2003ui}
  Balsara D S and Kim J S 2004
  {\it Astrophys. J.} {\bf 602} 1079

\bibitem{Balsara:2001rw}
  Balsara D 2001
  {\it J. Comput. Phys.} {\bf 174} 614

\bibitem{Dedner:2002}
Dedner A, Kemm F, Kr\"oner D, Munz C-D, Schnitzer T and Wesenberg M 2002
{\it J. Comput. Phys.} {\bf 175} 645

\bibitem{Brodbeck:1998az}
  Brodbeck O, Frittelli S, Hubner P and Reula O A 1999
  {\it J. Math. Phys.}  {\bf 40} 909

\bibitem{LiuOsher}
Liu X-D and Osher S 1998
  {\it J. Comput. Phys.} {\bf 142} 304


\bibitem{DelZanna:2002qr}
  Del Zanna L and Bucciantini N 2002
  {\it Astron. Astrophys.} {\bf 390} 1177

\bibitem{Lucas-Serrano:2004aq}
  Lucas-Serrano A, Font J A, Ibanez J M and Marti J M 2004
  Assessment of a high-resolution central scheme for the solution of the
  relativistic hydrodynamics equations
  {\it Preprint} astro-ph/0407541

\bibitem{Shibata:2005jv}
  Shibata M and Font J A 2005
  \PR D {\bf 72} 047501


\bibitem{DelZanna:2002rv}
  Del Zanna L, Bucciantini N and Londrillo P 2003
  {\it Astron. Astrophys.} {\bf 400} 397

\bibitem{Thornburg:2000cb}
  Thornburg J 2000
  A multiple-grid-patch evolution scheme for 3-D black hole excision
  {\it Preprint} gr-qc/0012012

\bibitem{Thornburg:2003sf}
  Thornburg J 2003
  {\it AIP Conf. Proc.}  {\bf 686} 247

\bibitem{Thornburg:2004dv}
  Thornburg J 2004
  \CQG {\bf 21} 3665

\bibitem{SebastianShu}
Sebastian K and Shu C-W 2003
  {\it SIAM J. Sci. Comput.} {\bf 19} 405

\bibitem{TangZhou}
  Tang H S  and Zhou T,
  {\it SIAM J. Numer. Anal.} {\bf 37} 173

\bibitem{ShuOsherI} 
Shu C-W and Osher S 1988
  {\it J. Comput. Phys.} {\bf 77} 439

\bibitem{ShuOsherII}
Shu C-W and Osher S 1989
  {\it J. Comput. Phys.} {\bf 83} 32

\bibitem{vanPutten}
 van Putten M H P M 2002
 {\it J. Math. Phys.} {\bf 43} 6195


\bibitem{Brackbill}
  Brackbill J U and Barnes D C 1980
  {\it J. Comp. Phys.} {\bf 35} 426

\bibitem{Komissarov:2002mp}
  Komissarov S S 2002
  Test problems for relativistic magnetohydrodynamics
  {\it Preprint} astro-ph/0209213

\bibitem{Komissarov:1999}
  Komissarov S S 1999
  {\it Mon. Not. Roy. Astron. Soc.} {\bf 303} 343


\bibitem{Giacomazzo:2005jy}
  Giacomazzo B and Rezzolla L 2005
  The Exact Solution of the Riemann Problem in Relativistic MHD
  {\it Preprint} gr-qc/0507102

\bibitem{Liu2005}
  Liu Y 2005
  {\it J. Comp. Phys.} {\bf 209} 82

\end{thebibliography}
\end{document}